\documentclass[12pt]{article}
\usepackage[utf8]{inputenc}
\usepackage{colortbl}
\usepackage{pifont}
\usepackage{url}
\usepackage{caption}
\usepackage{tabularx}
\usepackage{amssymb}
\usepackage{subcaption}
\usepackage[title]{appendix}
\usepackage{etoolbox}
\usepackage{algorithm}
\usepackage{algorithmicx}
\usepackage{siunitx}
\usepackage{setspace}
\usepackage{xcolor}
\makeatletter
\newcommand{\algocaption}[1]{\caption{#1}\let\footnote\oldfootnote}
\makeatother
\usepackage{subcaption}
\usepackage{amsthm}
\theoremstyle{definition}
\newtheorem{assumption}{Assumption}
\newtheorem{theorem}{Theorem}  
  
\newtheorem{proposition}{Proposition}
\usepackage{pdfpages}
\usepackage{float}
\usepackage{enumitem}

\usepackage[english]{babel}
\usepackage{bm}
\usepackage{diagbox}
\usepackage{natbib}
\usepackage{multirow}
\setcitestyle{aysep={}}
\usepackage{abstract}
\newcommand{\indentsmall}[1]{\hspace{2em}#1}

\usepackage[letterpaper,top=2cm,bottom=2cm,left=3cm,right=3cm,marginparwidth=1.75cm]{geometry}

\usepackage[hang]{footmisc}
\usepackage{amsmath}
\usepackage{graphicx}
\usepackage{tikz}

\usepackage[colorlinks=true, allcolors=blue]{hyperref}
\usepackage{cleveref}
\allowdisplaybreaks[4]
\title{Portfolio Optimization via Transfer Learning}
\date{ }
\newcommand{\covertitle}[1]{%
    {\Large\bfseries\centering #1 \par}
}

\begin{document}
\thispagestyle{empty} 
\vspace*{\fill}
\begin{center}
    \covertitle{Portfolio Optimization via Transfer Learning}
    \vspace{1cm}
    by\\
Kexin Wang\textsuperscript{1}, Xiaomeng Zhang\textsuperscript{2}, and Xinyu Zhang\textsuperscript{3a,3b}
\end{center}
\vspace*{\fill} 
\noindent 
\begin{minipage}[b][0pt]{\textwidth} 
\footnotesize
\hrule
\vspace{5pt}
\textsuperscript{1} School of Management, University of Science and Technology of China, Anhui, China, wkx1220@mail.ustc.edu.cn \\
\textsuperscript{2} Econometric Institute, Erasmus University Rotterdam, Burgemeester Oudlaan 50, 3062 PA Rotterdam, Netherlands, zhang@ese.eur.nl\\
\textsuperscript{3a} Academy of Mathematics and Systems Science, Chinese Academy of Sciences, Beijing, China, xinyu@amss.ac.cn\\
\textsuperscript{3b} School of Management, University of Science and Technology of China, Anhui, China, xinyu143@ustc.edu.cn\\
The authors would like to thank Professor Dacheng Xiu and Dr. Shiwei Huang for their insightful comments which have helped improve the manuscript substantially.
\end{minipage}
\setcounter{page}{1}
\pagenumbering{arabic} 
\maketitle

\begin{abstract}

Recognizing that asset markets generally exhibit shared informational characteristics, we develop a portfolio strategy based on transfer learning that leverages cross-market information to enhance the investment performance in the market of interest by forward validation. Our strategy asymptotically identifies and utilizes the informative datasets, selectively incorporating valid information while discarding the misleading information. This enables our strategy to achieve the maximum Sharpe ratio asymptotically. The promising performance is demonstrated by numerical studies and case studies of two portfolios: one consisting of stocks dual-listed in A-shares and H-shares, and another comprising equities from various industries of the United States.
\end{abstract}
\newpage
\doublespacing

\noindent Finding optimal portfolio strategies is a fundamental challenge for investors. 
When determining the portfolio strategies on a specific market or asset class that we are interested in,
the traditional strategies, such as the mean-variance portfolio theory proposed by \citet{r1} and its variants \citep{r10,r3,r5,r6,r12,r4,r7,ledoit2017,ao2018,r16}, are usually built up with the information solely on that specific market or asset class itself. 
However, data specific to the market or asset class  that we are interested in are often insufficient due to high volatility, structural breaks, and high-dimensional features, which can lead to suboptimal portfolio decisions. Moreover, when it comes to a emerging market or a newly thematic sector (e.g., ESG or crypto), its historical data is usually quite limited, making it more difficult to construct a reliable portfolio allocation.
Thus, this article is motivated to compensate for the insufficiency of the market dataset of interest.  

In this paper, we address the insufficiency of the market dataset of interest by transferring information from other related market datasets, such as those where the same stocks are traded. We propose a new portfolio strategy based on transfer learning technology.
It possesses the asymptotic optimality in terms of maximizing the Sharpe ratio index in the large sample sense, which can balance the expected return of investors with the risk they are taking on; see \cite{r17} for a review of the development of the Sharpe ratio.

Transfer learning, a branch of machine learning, focuses on transferring related knowledge from source domains (e.g., other related market datasets) to a potentially different target domain (e.g., the market dataset of interest);  see \cite{pan2010}, \cite{zhuang2011} and \cite{koshiyama2022} for the literature review.
The transfer learning technology is feasible in portfolio optimization due to the fact that many markets across sectors, industries, and asset classes often display shared patterns and inter-dependencies driven by macroeconomic factors, supply chain dynamics, investor behavior and so on \citep[see, e.g.,][]{buccheri2021,kelly2023,guo2025}. For example, the analysis of the H-shares and A-shares benefits significantly from incorporating information from the other market.  This utility stems from the fundamental commonality of cross-listed companies.  Since the corresponding H-shares and A-shares represent ownership in the same entity, their intrinsic value drivers, such as profitability and operational risk, are closely linked.  Consequently, price deviations in one market, often resulting from information asymmetry or local sentiment, can be identified through price movements in the other.  This spillover effect of information provides critical incremental insights for accurate asset valuation, mispricing detection, and the development of more robust investment strategies.

To transfer knowledge from other related markets for the portfolio optimization in the market of interest,
\citet{cao2023ssrn} developed a transfer learning technique, provided theoretical analyses of the transfer risk and applied it in three numerical experiments, including cross-continent transfer, cross-sector transfer and cross-frequency transfer.
However, they limited their strategy to transferring portfolio strategies from a single source dataset, rather than from multiple sources. 
Although the proposed transfer risk in this research can serve as a prior to figure out the suitable source datasets, relying on only a single source dataset may yield inferior results compared to leveraging multiple sources collectively when these source datasets are regarded as having relatively low risk. Moreover, estimating this transfer risk still requires additional assumptions about the distributions of the source and target data, such as assuming multivariate normality. By contrast, our strategy allows information to be transferred from multiple source datasets simultaneously and offers corresponding theoretical foundations.

Except for \citet{cao2023ssrn}, how to transfer information for portfolio construction remains relatively underexplored,  although transfer learning has been widely adopted within the financial domain to improve precision and efficiency in areas such as asset pricing \citep{r28,babi2023,lasse2024}, default risk assessment \citep{r29}, and modeling industrial chain interactions \citep{r30}. 
In the study of financial investment strategies, \citet{jeong2019} investigated trading system and involved transfer learning to handle the overfitting problem arising from small and highly volatile financial data. \citet{koshiyama2022} developed a fully end-to-end global trading architecture (QuantNet) based on transfer learning and meta-learning. By integrating diverse market data, QuantNet learns both market-agnostic trends and market-specific strategies by separating shared and specialized parameters, capturing general transferable dynamics and achieving superior returns. \citet{morstedt2024} focused on the
global minimum variance portfolio \citep{r1} and proposed a nonlinear shrinkage estimator for covariance estimation that determines the shrinkage parameters using cross-validation-based transfer learning. Specifically, cross-validation-based transfer learning requires that the validation set is formed from a disjoint source dataset of assets, disjoint from the target dataset.

In this article, we directly transfer portfolio strategies across different source datasets and select the transferring weight vector by forward validation. In addition, we do not need to assume that markets share specific parameters or structural components, which allows our strategy to remain flexible and agnostic to the underlying similarity structure across markets. 
We also provide three theoretical guarantees for our strategy. (i) Our strategy automatically excludes uninformative and misleading source datasets,  asymptotically concentrating portfolio weights on the informative datasets,
in contrast to \citet{cao2023ssrn}, which relies on pre-estimating transfer risk to choose the source dataset. (ii) Our portfolio strategy can asymptotically achieve the maximum attainable Sharpe ratio. (iii) The variance of the Sharpe ratio under our strategy is asymptotically lower than that of the $Non-transfer$ baseline strategy which only relies on the target dataset if there exists informative source dataset for the target market dataset. 
If none of the source dataset is informative for the target,  then the variance of the Sharpe ratio under our strategy is asymptotically equivalent to that of the $Non-transfer$ strategy. 

Our strategy is inspired by the optimal model averaging strategy \citep{hansen2007}, which is a frequentist model averaging technique aiming at optimally combine predictions or estimates from multiple candidate models; see \citet{enrique2015} for a review of the literature. More recently, \citet{hu2023} and \citet{zhang2024} have integrated the basic idea of optimal model averaging with transfer learning, introducing the concept of optimal transfer learning. However, both \citet{hu2023} and \citet{zhang2024} focused primarily on improving predictive accuracy and were not applicable for portfolio optimization. Hence, in this article, we adapt optimal transfer learning specifically for portfolio optimization. 
Although our strategy is developed in the context of portfolio optimization, the central idea can be readily extended to other decision-making problems. 

To validate the superiority of our strategy, we conduct numerical simulations and empirical applications focused on two scenarios: investing in stocks dual-listed in both H-shares and A-shares, and investing across different industries of the United States. Results confirm our theoretical conclusions that the inclusion of auxiliary datasets through transfer learning significantly enhances portfolio performance. 

The remainder of this article is organized as follows. \hyperref[sec:1]{Section 1} presents the problem of interest and the model framework. \hyperref[sec:2]{Section 2} establishes the asymptotic properties of our procedure. The performance of the proposed strategy is investigated via
numerical simulations in \hyperref[sec:3]{Section 3} and the empirical applications in \hyperref[sec:4]{Section 4}. \hyperref[sec:5]{Section 5} concludes the article with some discussion. All technical details including detailed proofs are provided in the Appendix.

\section{Model framework}\label{sec:1}

The concept of Sharpe ratio $(SR)$, also known as the Sharpe Index, was introduced by \citet{sharpe1966} to define a measure of the ratio of the return and the volatility. Its mathematical expression is
$$SR=\frac{\operatorname{E}(r_p)-r_f}{\sigma_p},$$ where $\operatorname{E}(r_p)$ represents the expected return of the portfolio, $r_f$ stands for risk-free asset and $\sigma_p$ is the standard deviation of the portfolio which is used to measure the overall risk of the portfolio. Obviously, Sharpe ratio not only focuses on the return of the assets, but also focuses on the risk of the assets.

Consider a capital market consisting of $d$ assets with the excess return vector $\bm{r}=(r_1,\dots,r_d)^{\top}$. If the corresponding asset allocation is $\bm{\phi}=(\phi_1,\dots,\phi_d)^{\top}$, the payoff is $\bm{\phi}^{\top}\bm{r}$. Denote the mean vector of the excess return as $\bm{\mu}$, the covariance matrix as $\bm{\Sigma}$ and $\Phi=\{\bm{\phi} \in [0, 1]^ d|\  \sum_{i=1}^d\phi_i = 1 \}$. The optimal portfolio problem is to find the maximum Sharpe ratio by solving the following optimization problem.\begin{equation*}
\bm{\phi}_o =\underset{\bm{\phi} \in \Phi }{\arg \max }\frac{\operatorname{E}\left(\bm{\phi}^{\top} \bm{r}\right)}{\sqrt{\operatorname{Var}\left(\bm{\phi}^{\top} \bm{r}\right)}}=\underset{\bm{\phi} \in \Phi}{\arg \max } \frac{\bm{\mu}^{\top} \bm{\phi}}{\sqrt{\bm{\phi}^{\top} \bm{\Sigma} \bm{\phi}}} ,\tag{1}\label{1}
\end{equation*}
where $\bm{\phi}_o$ represents the optimal portfolio allocation that maximizes the Sharpe ratio. 

Since there may exist other available assets which provide useful information when investing in the target assets, we want to make full use of the possible effective information to invest in the target assets. Assume that there are $M$ available source datasets, and denote the target data as $T$ and the source data as $\{S_m|\ m=1,\dots,M\}$, respectively. We treat $M$ as fixed and allow $d$ to diverge. The sample sizes of the target data $T$ and the source data $S_m$ are $N_0$ and $N_m$, respectively. Let $\widetilde{N}=\text{min}\{N_0,\dots,N_M\}$. For simplicity, denote the excess return of the target assets as $\{\bm{r}_{0,t}\in \text{R}^d|\ t=-(N_0-\widetilde{N}-1),\dots,\widetilde{N}\}$ and the excess return of the source assets as $\{\bm{r}_{m,t}\in \text{R}^d|\ m=1,\dots,M,\ t=-(N_m-\widetilde{N}-1),\dots,\widetilde{N}\}$, respectively. For $m=0,1,\dots,M$ and $t=-(N_m-\widetilde{N}-1),\dots,\widetilde{N}$, denote the mean vector of the excess return $\bm{r}_{m,t}$ as $\bm{\mu}_{m,t}$ and the covariance matrix of the excess return $\bm{r}_{m,t}$ as $\bm{\Sigma}_{m,t}$. Under these true parameters, the optimal asset allocation, denoted as $\bm{\phi}_{m,t}$, can be derived using \eqref{2}.
\begin{equation*}
   \bm{\phi}_{m,t} =\underset{\bm{\phi} \in \Phi}{\arg \max } \frac{\bm{\mu}_{m,t}^{\top} \bm{\phi}}{\sqrt{\bm{\phi}^{\top} \bm{\Sigma}_{m,t} \bm{\phi}}} .\tag{2}\label{2} 
\end{equation*} However, we never know these parameters in practice. Empirically, $\bm{\mu}_{m,t}$ and $\bm{\Sigma}_{m,t}$ are estimated from the historical return. For $m=0,1,\dots,M,\ t=2,\dots,\widetilde{N}$, we use the historical sample mean and historical sample covariance to estimate. That is, the estimators $\widehat{\bm{\mu}}_{m,t}$ and $\widehat{\bm{\Sigma}}_{m,t}$ for $\bm{\mu}_{m,t}$ and $\bm{\Sigma}_{m,t}$ can be calculated using\begin{equation*}
\widehat{\bm{\mu}}_{m,t}=\frac{1}{t-1+N_m-\widetilde{N}}\sum_{j=-(N_m-\widetilde{N}-1)}^{t-1} \bm{r}_{m,j} \tag{3}\label{3}
\end{equation*}and
\begin{equation*}
\widehat{\bm{\Sigma}}_{m,t}=\frac{1}{t-2+N_m-\widetilde{N}}\sum_{j=-(N_m-\widetilde{N}-1)}^{t-1}  (\bm{r}_{m,j}-\widehat{\bm{\mu}}_{m,t})(\bm{r}_{m,j}-\widehat{\bm{\mu}}_{m,t})^{\top},\tag{4}\label{4}
\end{equation*}
respectively. Given $\widehat{\bm{\mu}}_{m,t}$ and $\widehat{\bm{\Sigma}}_{m,t}$, the optimal investment strategy can be estimated.\begin{equation*}
\widehat{\bm{\phi}}_{m,t}=\underset{\bm{\phi} \in \Phi}{\arg \max } \frac{\widehat{\bm{\mu}}_{m,t}^{\top} \bm{\phi}}{\sqrt{\bm{\phi}^{\top} \widehat{\bm{\Sigma}}_{m,t} \bm{\phi}}} .\tag{5}\label{5}
\end{equation*} Denote the set $\mathcal{W}=\{\bm{w}\in [0,1]^{M+1}|\ \sum_{m=0}^{M}\ w_m=1\}$ and the weight vector $\bm{w}=(w_0,w_1,\dots,w_M)^{\top}$. Define the weighted allocation of the target data at time $t$ as
\begin{equation*}
\widehat{\bm{\phi}}_t(\bm{w})=\sum_{m=0}^M w_m \widehat{\bm{\phi}}_{m,t}.\tag{6}\label{6}
\end{equation*}
As specified in \eqref{6}, the proposed estimator combines optimal portfolio estimators from multiple source datasets at time $t$ through a weighted integration scheme, where the weight vector $\bm{w}$ facilitates efficient information transfer across domains.

\subsection{A motivation example}\label{sec:1.1}

To demonstrate the advantages of incorporating source assets, we construct a simplified example where all datasets span $\widetilde{N}$ identical time periods. For analytical tractability, we assume that the excess returns $\{\bm{r}_{m,t}|\ t=1,\dots,\widetilde{N}\}$ are independently and identically distributed both across different datasets and within each individual dataset across $t$. For each dataset, calculate $\widehat{\bm{\mu}}_{m,\widetilde{N}+1}$ and $\widehat{\bm{\Sigma}}_{m,\widetilde{N}+1}$ according to \eqref{3} and \eqref{4}, respectively. Thus, applying \eqref{5}, we can obtain $\widehat{\bm{\phi}}_{0,\widetilde{N}+1}$ for the target assets $T$ and $\{\widehat{\bm{\phi}}_{m,\widetilde{N}+1}|\ m=1,\dots,M\}$ for the source assets $\{S_m|\ m=1,\dots,M\}$, respectively. Denote $\widehat{\bm{\phi}}^{equal}_{\widetilde{N}+1}$ as the equal weighted allocation $\sum_{m=0}^M\widehat{\bm{\phi}}_{m,\widetilde{N}+1}/(M+1)$. A direct performance comparison is made through Sharpe ratio $\widetilde{SR}_{0,\widetilde{N}+1}$ and $\widetilde{SR}_{0,\widetilde{N}+1}^{equal}$ evaluation at time $\widetilde{N}+1$, where $$\widetilde{SR}_{0,\widetilde{N}+1}=\frac{\bm{\mu}_{0,\widetilde{N}+1}^{\top} \widehat{\bm{\phi}}_{0,\widetilde{N}+1}}{\sqrt{\widehat{\bm{\phi}}_{0,\widetilde{N}+1}^{\top} \bm{\Sigma}_{0,\widetilde{N}+1} \widehat{\bm{\phi}}_{0,\widetilde{N}+1}}},\ \widetilde{SR}_{0,\widetilde{N}+1}^{equal}=\frac{\bm{\mu}_{0,\widetilde{N}+1}^{\top} \widehat{\bm{\phi}}_{\widetilde{N}+1}^{equal}}{\sqrt{(\widehat{\bm{\phi}}_{\widetilde{N}+1}^{equal})^{\top} \bm{\Sigma}_{0,\widetilde{N}+1} \widehat{\bm{\phi}}_{\widetilde{N}+1}^{equal}}}.$$ Denote $SR_{0,\widetilde{N}+1}$ as the maximum attainable Sharpe ratio for the target assets.
$$SR_{0,\widetilde{N}+1}=\frac{\bm{\mu}_{0,\widetilde{N}+1}^{\top} \bm{\phi}_{0,\widetilde{N}+1}}{\sqrt{\bm{\phi}_{0,\widetilde{N}+1}^{\top} \bm{\Sigma}_{0,\widetilde{N}+1} \bm{\phi}_{0,\widetilde{N}+1}}}.$$
\begin{proposition}\label{prop1} When $\widetilde{N}\rightarrow \infty$, if $\underset{\left\|\bm{a}\right\|_2=1,\ \bm{a}\in \text{R}^d}{\text{sup}}\text{E}\left\| \bm{a}^{\top}(\bm{r}_{m,t}-\bm{\mu}_{m,t})\right\| ^4\leq c_1<\infty$ and $0<c_2\leq \lambda_{min}(\Sigma_{0,\widetilde{N}+1})$ hold uniformly for some positive constants $c_1$ and $c_2$, and $d/\widetilde{N}\rightarrow0$, then
$$\frac{\widetilde{SR}_{0,\widetilde{N}+1}}{SR_{0,\widetilde{N}+1}}=1+o_p(1),\ \ \frac{\widetilde{SR}_{0,\widetilde{N}+1}^{equal}}{SR_{0,\widetilde{N}+1}}=1+o_p(1),$$
$$\frac{\text{Var}(\widetilde{SR}^{equal}_{0,\widetilde{N}+1})}{\text{Var}(\widetilde{SR}_{0,\widetilde{N}+1})}\rightarrow \frac{1}{(M+1)^2}.$$
\end{proposition}

\hyperref[prop1]{Proposition 1} implies that the resulting Sharpe ratio will converge to the maximum value $SR_{0,\widetilde{N}+1}$ in probability, irrespective of whether we employ $\widehat{\bm{\phi}}_{0,\widetilde{N}+1}$ or $\widehat{\bm{\phi}}^{equal}_{\widetilde{N}+1}$. However, the asymptotic variance differs significantly between these strategies. Specifically, $\widetilde{SR}_{0,\widetilde{N}+1}^{equal}$ achieves a superior convergence rate, as it exhibits a lower variance. See Appendix E for the proof of \hyperref[prop1]{Proposition 1}. The following example provides an intuitive illustration of \hyperref[prop1]{Proposition 1}. 

\noindent\textbf{Example 1.} 
Let $M = 5$ and assume the asset returns of the source and target assets follow the same multivariate normal distribution $\text{MVN}(\bm{\mu},\bm{\Sigma})$. That is, for $m=0,1,\dots,5$ and $t=1,\dots,\widetilde{N}$,
\begin{equation*}
    \bm{r}_{m,t}\sim \text{MVN}(\bm{\mu},\bm{\Sigma}),\tag{7}\label{7}
\end{equation*}
where $\bm{\mu}=(1.5,1.9,2.8,1.7,-0.9)^{\top}$ and the components of the covariance matrix are
$\bm{\Sigma}(i,j)=0.5^{|i-j|}$. Generate the target data and the source data with the sample size $\widetilde{N}$ using \eqref{7},  respectively, where $\widetilde{N}\in\{30,60,90,120,150,250,300,400,500\}$. Calculate $\widetilde{SR}_{0,\widetilde{N}+1}$ and $\widetilde{SR}^{equal}_{0,\widetilde{N}+1}$. Let $ESR_{0,\widetilde{N}+1}$ and $ESR^{equal}_{0,\widetilde{N}+1}$ denote the sample mean of $\widetilde{SR}_{0,\widetilde{N}+1}$ and $\widetilde{SR}^{equal}_{0,\widetilde{N}+1}$ over $1000$ replications, respectively. Let $VSR_{0,\widetilde{N}+1}$ and $VSR^{equal}_{0,\widetilde{N}+1}$ denote the sample variance of $\widetilde{SR}_{0,\widetilde{N}+1}$ and $\widetilde{SR}^{equal}_{0,\widetilde{N}+1}$ over $1000$ replications, respectively. In detail, the four indexes can be derived using the following formulas. Denote $\widehat{\bm{\phi}}^j_{0,\widetilde{N}+1}$ and $\widehat{\bm{\phi}}_{\widetilde{N}+1}^{equal,j}$ as the asset allocation calculated in the $j$th replication, respectively.
\begin{equation*}\begin{aligned}
\widetilde{SR}_{0,\widetilde{N}+1}^j=\frac{\bm{\mu}_{0,\widetilde{N}+1}^{\top} \widehat{\bm{\phi}}^j_{0,\widetilde{N}+1}}{\sqrt{\widehat{\bm{\phi}}_{0,\widetilde{N}+1}^{j\top} \bm{\Sigma}_{0,\widetilde{N}+1} \widehat{\bm{\phi}}^j_{0,\widetilde{N}+1}}},&\ \ \widetilde{SR}_{0,\widetilde{N}+1}^{equal,j}=\frac{\bm{\mu}_{0,\widetilde{N}+1}^{\top} \widehat{\bm{\phi}}_{\widetilde{N}+1}^{equal,j}}{\sqrt{(\widehat{\bm{\phi}}_{\widetilde{N}+1}^{equal,j})^{\top} \bm{\Sigma}_{0,\widetilde{N}+1} \widehat{\bm{\phi}}_{\widetilde{N}+1}^{equal,j}}},\\
ESR_{0,\widetilde{N}+1}=\frac{1}{1000}\sum_{j=1}^{1000}\widetilde{SR}_{0,\widetilde{N}+1}^j,&\ \ ESR^{equal}_{0,\widetilde{N}+1}=\frac{1}{1000}\sum_{j=1}^{1000}\widetilde{SR}^{equal,j}_{0,\widetilde{N}+1},\\
VSR_{0,\widetilde{N}+1}=\frac{1}{1000}\sum_{j=1}^{1000}&(\widetilde{SR}_{0,\widetilde{N}+1}^j-ESR_{0,\widetilde{N}+1})^2,\\
VSR^{equal}_{0,\widetilde{N}+1}=\frac{1}{1000}\sum_{j=1}^{1000}&(\widetilde{SR}^{equal,j}_{0,\widetilde{N}+1}-ESR^{equal}_{0,\widetilde{N}+1})^2.
\end{aligned}\end{equation*}
Furthermore, to evaluate the relative performance of the equal-weighted allocation strategy, we conduct a comparative analysis with a baseline strategy that augments the target dataset size. Specifically, we compute the portfolio $\widehat{\bm{\phi}}_{0,\widetilde{N}+1}^{Pool}$ by pooling all available observations from each dataset.
We proceed to compute the resulting Sharpe ratio, denoted as $\widetilde{SR}_{0,\widetilde{N}+1}^{Pool}$, for the merged-sample strategy. Let $\widehat{\bm{\phi}}_{0,\widetilde{N}+1}^{Pool,j}$ and $\widetilde{SR}_{0,\widetilde{N}+1}^{Pool,j}$ denote the portfolio strategy and the resulting Sharpe ratio estimated in the $j$th replication by data merging, respectively. The expected Sharpe ratio $ESR_{0,\widetilde{N}+1}^{Pool}$ is estimated using the sample average over $1000$ independent replications. That is,
\begin{equation*}
ESR^{Pool}_{0,\widetilde{N}+1}=\frac{1}{1000}\sum_{j=1}^{1000}\widetilde{SR}^{Pool,j}_{0,\widetilde{N}+1},\ \text{where}\  \widetilde{SR}_{0,\widetilde{N}+1}^{Pool,j}=\frac{\bm{\mu}_{0,\widetilde{N}+1}^{\top} \widehat{\bm{\phi}}_{\widetilde{N}+1}^{Pool,j}}{\sqrt{(\widehat{\bm{\phi}}_{\widetilde{N}+1}^{Pool,j})^{\top} \bm{\Sigma}_{0,\widetilde{N}+1} \widehat{\bm{\phi}}_{\widetilde{N}+1}^{Pool,j}}}.
\end{equation*}

\begin{figure}[h] 
    \centering 
    
    \begin{minipage}[t]{0.49\textwidth}
        \centering
        \subcaption{Simulation comparison of Sharpe ratio}
        \includegraphics[width=\linewidth]{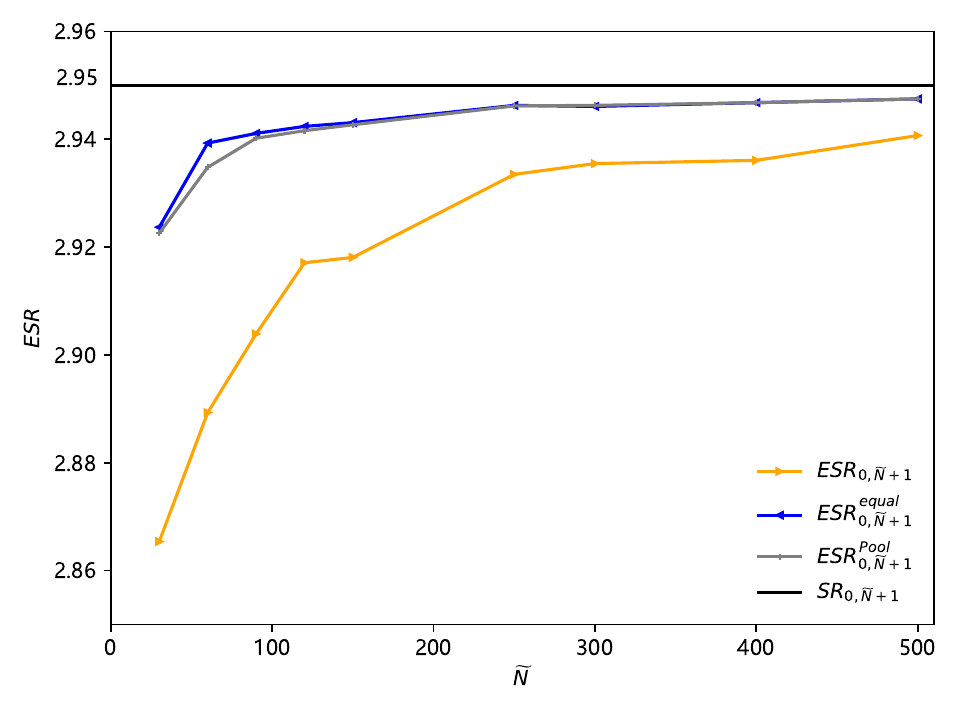}
        \label{SRme}
    \end{minipage}
    \begin{minipage}[t]{0.49\textwidth}
        \centering
        \subcaption{Simulation comparison of variance}
        \includegraphics[width=\linewidth]{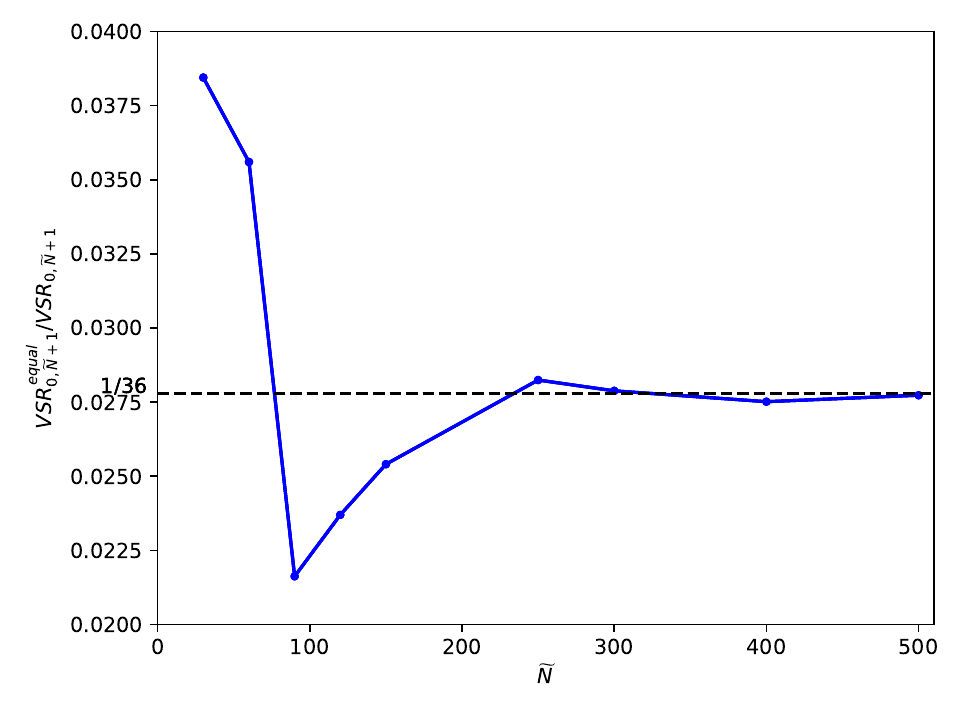}
        \label{VRme}
    \end{minipage}
    \captionsetup{singlelinecheck=off, justification=raggedright}
    \caption{Performance of different portfolio strategies}
  \raggedright\footnotesize
  The figure presents a comparative analysis of the Sharpe ratio values obtained through different portfolio strategies across varying sample sizes $\widetilde{N}$. In the left panel, we demonstrate $ESR_{0,\widetilde{N}+1}^{equal}$ (blue line with left-pointing marks), $ESR_{0,\widetilde{N}+1}^{Pool}$ (gray line with star markers), $ESR_{0,\widetilde{N}+1}$ (orange line with right-pointing markers) and the maximum Sharpe ratio attainable for the target (black line). In the right panel, we demonstrate the sample $VSR_{0,\widetilde{N}+1}^{equal}/VSR_{0,\widetilde{N}+1}$ (blue line with circular markers) and the theoretical limit of the ratio (black line).
    \label{me}
\end{figure}

With the population mean vector $\bm{\mu}$ and covariance matrix $\bm{\Sigma}$ specified, the attainable maximum Sharpe ratio is analytically determined to be $SR_{0,\widetilde{N}+1}=2.95$ as our benchmark. \hyperref[me]{Figure 1} demonstrates that both $\widetilde{SR}_{0,\widetilde{N}+1}$ and $\widetilde{SR}^{equal}_{0,\widetilde{N}+1}$ exhibit asymptotic convergence to the theoretical maximum $SR_{0,\widetilde{N}+1}$. However, $\widetilde{SR}^{equal}_{0,\widetilde{N}+1}$ exhibits superior finite sample properties, consistently dominating $\widetilde{SR}_{0,\widetilde{N}+1}$ across all $\widetilde{N}$. Under the identical data generating processes of each dataset, incorporating additional source data effectively expands the informational basis for estimation. The analysis of the variance indicates that as $\widetilde{N} \to \infty$, $VSR_{0,\widetilde{N}+1}^{equal}/VSR_{0,\widetilde{N}+1}$ converges to its theoretical limit $1/(M+1)^2=1/36$. This also indicates that $\widetilde{SR}^{equal}_{0,\widetilde{N}+1}$ exhibits a higher convergence rate to its asymptotic limit compared to the conventional estimator $\widetilde{SR}_{0,\widetilde{N}+1}$.

\subsection{Transfer learning strategy by forward validation}\label{sec:1.2}
In the motivation example, we demonstrate the efficacy of integrating data from multiple sources to enhance the investment performance of the target. Our findings reveal that even a naive equal weighted combination strategy yields significant improvements. Nevertheless, empirical evidence often reveals substantial heterogeneity across different markets, which renders equal weighted combination suboptimal and may induce negative transfer effects that significantly degrade model performance. To better accommodate the potential heterogeneity in empirical data, we propose a novel data-driven weighting strategy that adaptively adjusts to market conditions and improves the investment performance of the target market at time $\widetilde{N}+1$. To choose the weight $\bm{w}$ in $\widehat{\bm{\phi}}_t(\bm{w})$, we propose the criterion in \hyperref[alg1]{Algorithm 1}. Denote $[x]$ as the maximum integer that do not larger than $x$. 
\begin{algorithm}[H]
    \caption{Transfer Learning ($TL$) algorithm}
    \label{alg1}
    \begin{algorithmic}[0]
        \State\textbf{Step 1. Split the target data and source data.}\label{step1}\\
            \indentsmall{Divide each dataset into $[\widetilde{N}/h]$ disjoint parts in time order. Within this partitioning, the initial segment of each dataset contains $N_m-([\widetilde{N}/h]-1)h$ samples, while all subsequent segments maintain a constant size of $h$ samples.}
        \State \textbf{Step 2. Renumbering the timestamp.}\label{step2}\\
            \indentsmall{After each interval boundary, index the subsequent temporal points as $\tau_i$ where $i=1,\dots,[\widetilde{N}/h]$.}\footnotemark
        \State \textbf{Step 3. Calculate the related parameter estimators of each dataset at time $\tau_i$.}\label{step3}\\
            \indentsmall{For $m=0$ to $M$:}
            \begin{enumerate}[leftmargin=2cm]
                \item Calculate $\widehat{\bm{\mu}}_{m,\tau_i}$ and $\widehat{\bm{\Sigma}}_{m,\tau_i}$ using the latest $h$ samples and \eqref{3}-\eqref{4}, respectively.
                \item Estimate $\bm{\mu}_{m,\tau_i}$ and $\bm{\Sigma}_{m,\tau_i}$ using all the former samples and \eqref{3}-\eqref{4}. Given the estimators, apply \eqref{5} to calculate $\widehat{\bm{\phi}}_{m,\tau_i}$.
            \end{enumerate}
        \State \textbf{Step 4. Solve weights.}\label{step4}\\
            \indentsmall{Solve \eqref{8} to obtain the optimal transferring weight.}
            \begin{equation*}
\widehat{\bm{w}}=\underset{\bm{w} \in \mathcal{W}}{\arg \max }\ 
\frac{1}{[\widetilde{N}/h]-1}\sum_{i=1}^{[\widetilde{N}/h]-1 }\frac{(\widehat{\bm{\mu}}_{0,\tau_{i+1}}^{\top}) \widehat{\bm{\phi}}_{\tau_i}(\bm{w})}{\sqrt{\widehat{\bm{\phi}}_{\tau_i}(\bm{w})^{\top}\widehat{\bm{\Sigma}}_{0,\tau_{i+1}}\widehat{\bm{\phi}}_{\tau_i}(\bm{w})}} .\tag{8}\label{8}
\end{equation*}
        \State \textbf{Step 5. Calculate the asset allocation at time $\widetilde{N}+1$.}\label{step5}\\
        \indentsmall{Let $\widehat{w}_m$ be the $(m+1)$th component of $\widehat{\bm{w}}$. Hence,}\begin{equation*}\widehat{\bm{\phi}}_{\widetilde{N}+1}(\widehat{\bm{w}})=\sum\limits_{m=0}\limits^M \widehat{w}_m \widehat{\bm{\phi}}_{m,\widetilde{N}+1}.\tag{9}\label{9}\end{equation*}
    \end{algorithmic}
\end{algorithm}
\footnotetext{ For pedagogical clarity, we demonstrate \hyperref[step1]{Step 1} and \hyperref[step2]{Step 2} through a specific case. Consider a scenario with parameters $M=3,\ N_0=300,\ N_1=400,\ N_2=500,\ N_3=600,\ h=50$. In this configuration, each dataset is partitioned into $[\widetilde{N}/h]=[N_0/h]=6$ parts and the subsequent evaluation time points after each interval boundary are $t\in\{51,101,151,201,251,301\}$. Relabel these time points as $\tau_i$, where $\ i=1,\dots,6$, respectively.}

\begin{figure}[H]
    \centering
\includegraphics[height=500pt]{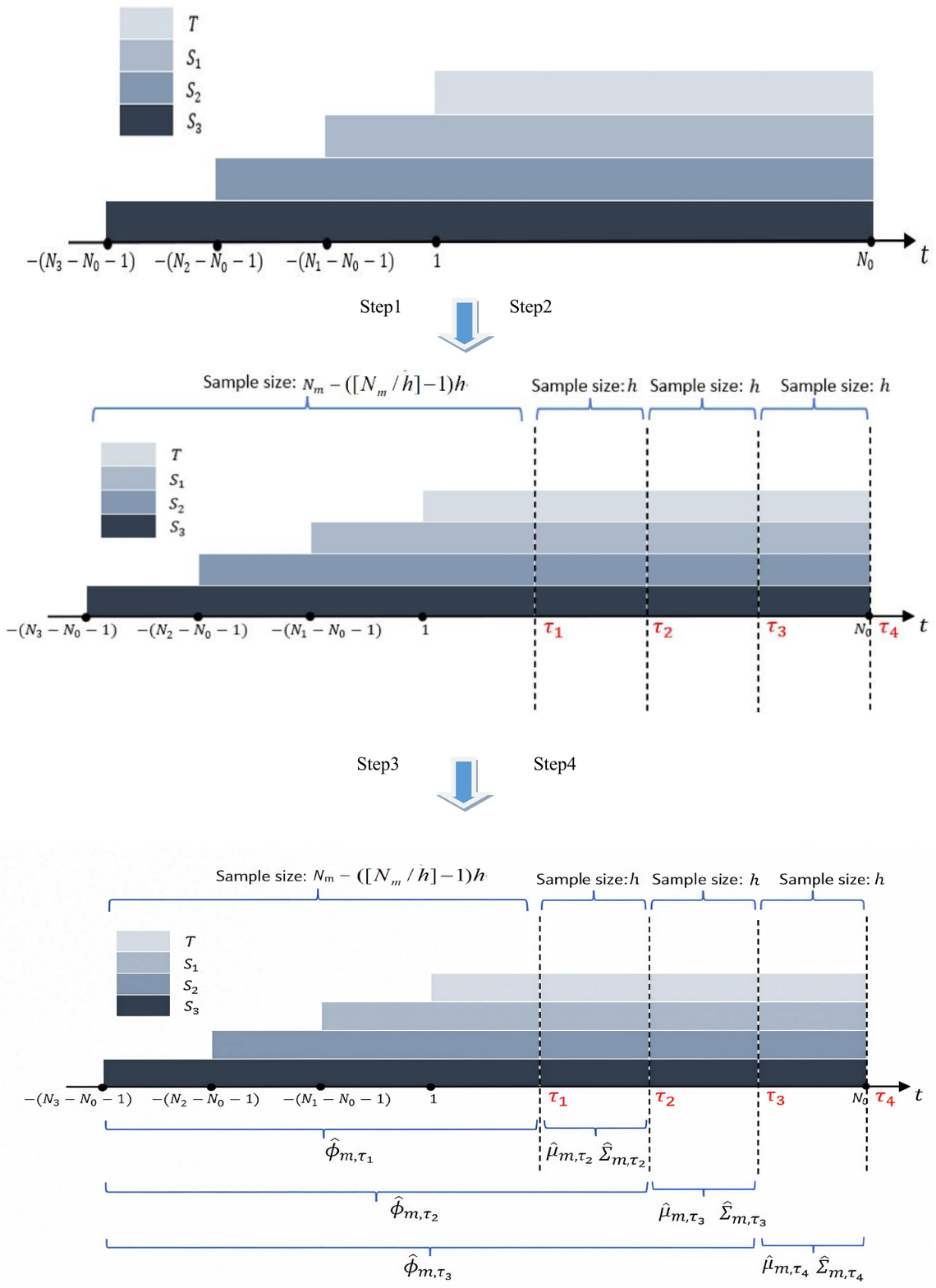}
\captionsetup{singlelinecheck=off, justification=raggedright}
    \caption{Schematic illustration of the proposed $TL$ strategy}
    \raggedright\footnotesize
    The figure provides an intuitive visualization of the operational workflow of the proposed $TL$ strategy for the case where $M=3$ and $\widetilde{N}=N_0$.
    \label{process}
\end{figure}

The schematic diagram in \hyperref[process]{Figure 2} provides an intuitive illustration of the algorithmic workflow. For demonstrative purposes and without loss of generality, we consider the case where $\widetilde{N}=N_0$ in this exposition.

\section{Asymptotic properties}\label{sec:2}
In this section, we demonstrate that our proposed $TL$ strategy can
automatically assign all weights to the informative datasets and asymptotically achieve the maximum attainable
Sharpe ratio. Furthermore, we conduct a comparative analysis of the variance between the Sharpe ratio achieved through our proposed $TL$ strategy and the $Non-transfer$ strategy. To present the asymptotic properties, we need some regularity conditions. Unless otherwise stated, all limiting properties are set when the sample size of each dataset goes to infinity.

\begin{assumption}\label{ass:a1}For $m=0,1,\dots,M$, there exist a constant $\zeta>0$, parameters $\{\bm{\mu}^*_{m,\tau_i}|\ i=1,\dots,[\widetilde{N}/h]\}$ and $\{\bm{\Sigma}^*_{m,\tau_i}|\ i=1,\dots,[\widetilde{N}/h]\}$, such that

$(\text{i})$\label{ass:a1i} $d^{\zeta}h^{-\frac{1}{2}}=o(1)$;

$(\text{ii})$\label{ass:a1ii} $ h^{\frac{1}{2}}d^{-\zeta}\parallel \widehat{\bm{\mu}}_{m,\tau_i}-\bm{\mu}^*_{m,\tau_i}\parallel_2=O_p(1)$ and $h^{\frac{1}{2}}d^{-\zeta}\parallel \widehat{\bm{\Sigma}}_{m,\tau_i}-\bm{\Sigma}^*_{m,\tau_i}\parallel_\text{F}=O_p(1);$

$(\text{iii})$\label{ass:a1iii} $h^{\frac{1}{2}}d^{-\zeta}\parallel \bm{\mu}^*_{m,\tau_{i+1}}-\bm{\mu}^*_{m,\tau_i}\parallel_2=O(1)$ and $h^{\frac{1}{2}}d^{-\zeta}\parallel \bm{\Sigma}^*_{m,\tau_{i+1}}-\bm{\Sigma}^*_{m,\tau_i}\parallel_\text{F}=O(1).$
\end{assumption}
\hyperref[ass:a1i]{Assumption 1(i)} restricts the divergence rate of $d$. \hyperref[ass:a1ii]{Assumption 1(ii)} requires that there exist convergence relationships between the historical sample estimators $\widehat{\bm{\mu}}_{m,\tau_i}$, $\widehat{\bm{\Sigma}}_{m,\tau_i}$ and the limit parameters $\bm{\mu}^*_{m,\tau_i}$, $\bm{\Sigma}^*_{m,\tau_i}$, respectively.
\hyperref[ass:a1iii]{Assumption 1(iii)} imposes additional conditions on the sequences $\{\bm{\mu}^*_{m,\tau_i}|\ i=1,\dots,[\widetilde{N}/h]\}$ and $\{\bm{\Sigma}^*_{m,\tau_i}|\ i=1,\dots,[\widetilde{N}/h]\}$ by requiring the internal convergence. In the case of multivariate sample mean and multivariate sample covariance, the reviews of the literature on the
dependence of the error of approximation on the dimension $d$ are given \citep[see, e.g.,][]{Zitikis2006,bentkus1986,bentkus2003,bloznelis1989,gotze1991,nagaev2006}. \cite{bentkus2003} proves the existence of $\zeta$, where $\zeta=1/4$, $\{\bm{\mu}^*_{m,\tau_i}|\ i=1,\dots,[\widetilde{N}/h]\}$ and $\{\bm{\Sigma}^*_{m,\tau_i}|\ i=1,\dots,[\widetilde{N}/h]\}$ under some specific conditions where the samples are $i.i.d$, the expectation exists, the covariance matrix is positive definite and the third-order moment is finite. This assumption remains valid according to the conclusion of \cite{Zitikis2006}, where the relationship between the convergence rates of the multivariate sample mean, multivariate sample covariance matrix and the divergent rate of the parameters $d$ and $h$ are established under other regularity conditions.

Denote 
$$\widetilde{SR}^*_{0,\tau_{i+1}}(\bm{w})= \frac{(\bm{\mu}_{0,\tau_{i+1}}^{\top}) \widehat{\bm{\phi}}_{\tau_{i}}(\bm{w})}{\sqrt{\widehat{\bm{\phi}}_{\tau_{i}}(\bm{w})^{\top}\bm{\Sigma}_{0,\tau_{i+1}}\widehat{\bm{\phi}}_{\tau_{i}}(\bm{w})}},\ \ \widetilde{MS}^*_{\widetilde{N}}(\bm{w})=\frac{1}{[\widetilde{N}/h]-1}\sum_{i=1}^{[\widetilde{N}/h]-1 }\widetilde{SR}^*_{0,\tau_{i+1}}(\bm{w}),$$ 
$$SR_{0,\tau_{i+1}}=\frac{(\bm{\mu}_{0,\tau_{i+1}}^{\top}) \bm{\phi}_{0,\tau_{i+1}}}{\sqrt{\bm{\phi}_{0,\tau_{i+1}}^{\top}\bm{\Sigma}_{0,\tau_{i+1}}\bm{\phi}_{0,\tau_{i+1}}}},\ \ MS_{\widetilde{N}}=\frac{1}{[\widetilde{N}/h]-1}\sum_{i=1}^{[\widetilde{N}/h]-1 }SR_{0,\tau_{i+1}}.$$

We can calculate the corresponding asset allocation $\bm{\phi}_{m,\tau_i}^*$ under the parameters $\bm{\mu}^*_{m,\tau_i}$, $\bm{\Sigma}^*_{m,\tau_i}$ and Equation \eqref{1}. If $\parallel\bm{\phi}_{0,\tau_i}-\bm{\phi}_{m,\tau_i}^*\parallel_2=o(1)$ holds for each $\tau_i$, define the corresponding dataset as an effective information set which can provide effective information for the investment in the target. Under \hyperref[ass:a1]{Assumption 1}, if the limit parameters $\bm{\mu}^*_{m,\tau_i}$ and $\bm{\Sigma}^*_{m,\tau_i}$ satisfy that $\parallel \bm{\mu}_{0,\tau_{i}}-\bm{\mu}^*_{m,\tau_i}\parallel_2=o(1)$ and $\parallel \bm{\Sigma}_{0,\tau_{i}}-\bm{\Sigma}^*_{m,\tau_i}\parallel_\text{F}=o(1)$, then $\parallel \bm{\phi}_{0,\tau_i}-\bm{\phi}_{m,\tau_i}^*\parallel_2=o(1)$, which implies that the dataset with index $m$ is an effective information set when the difference between the corresponding limit parameters $\bm{\mu}^*_{m,\tau_i}$, $\bm{\Sigma}^*_{m,\tau_i}$ and the population parameters of the target $\bm{\mu}_{0,\tau_i}$, $\bm{\Sigma}_{0,\tau_i}$ asymptotically converge to $0$, respectively. Let $\mathcal{D}\subseteq\{0,1,\dots,M\}$ denote the index set of datasets which can provide effective information for the target and $\mathcal{D}^c$ denote the
complement of $\mathcal{D}$. Denote $\Gamma(\bm{w}) =\sum_{m\in \mathcal{D}}w_m$, where $w_m$ is the $(m+1)$th element of $\bm{w}$, then $\Gamma(\widehat{\bm{w}}) =\sum_{m\in \mathcal{D}}\widehat{w}_m$. Let $\widetilde{\xi}^*_{\widetilde{N}}=MS_{\widetilde{N}}-\text{sup}_{\bm{w} \in \mathcal{W},\Gamma(\bm{w})=0}\ \widetilde{MS}^*_{\widetilde{N}}(\bm{w})$.

\begin{assumption}\label{ass:a2} $\widetilde{\xi}_{\widetilde{N}}^{*-1}h^{-\frac{1}{2}}d^{\zeta}=o(1)$.\end{assumption}
\hyperref[ass:a2]{Assumption 2} postulates the convergence rate of the difference between the maximum Sharpe ratio $MS_{\widetilde{N}}$ and the weighted  Sharpe ratio $\widetilde{MS}^*_{\widetilde{N}}(\bm{w})$ when the weight proportion on the datasets which can provide effective information is set to be $0$. If there exists a source dataset $S_j$ which just provides ineffective information, in other words, the corresponding optimal asset allocation under the parameters $\bm{\mu}^*_{m,\tau_i}$ and $\bm{\Sigma}^*_{m,\tau_i}$, denoted as $\bm{\phi}^*_{j,t}$, is greatly different from the optimal asset allocation $\bm{\phi}_{0,t}$, the Sharpe ratio obtained using the allocation $\bm{\phi}^*_{j,t}$ must be less than the maximum Sharpe ratio. From this point of view, this assumption is easy to be satisfied. 

\begin{assumption}\label{ass:a3}\ 

$(\text{i})$ $ h^{\frac{1}{2}}d^{-\zeta}\parallel \bm{\mu}_{0,\tau_{i}}-\bm{\mu}^*_{0,\tau_i}\parallel_2=O(1)$;

$(\text{ii})$ $h^{\frac{1}{2}}d^{-\zeta}\parallel \bm{\Sigma}_{0,\tau_{i}}-\bm{\Sigma}^*_{0,\tau_i}\parallel_\text{F}=O(1).$
\end{assumption}
\hyperref[ass:a3]{Assumption 3} imposes convergence constraints on parameters $\bm{\mu}^*_{0,\tau_i}$ and $\bm{\Sigma}^*_{0,\tau_i}$, which can be economically rationalized through two fundamental market dynamics.  First, the assumption captures the local stationarity property observed in high-frequency financial data \citep{diebold2001} through $\bm{\mu}^*_{0,\tau_i}$ and $\bm{\Sigma}^*_{0,\tau_i}$, where market microstructure effects dominate short-term price movements.  Second, it accommodates the equilibrium convergence behavior characteristic of long-term market data \citep{cochrane2009} through the convergence properties, where fundamental economic forces drive asset prices towards their steady-state values. Combined with \hyperref[ass:a1]{Assumption 1}, \hyperref[ass:a3]{Assumption 3} implies that the historical sample estimators $\widehat{\bm{\mu}}_{0,\tau_i}$ and $\widehat{\bm{\Sigma}}_{0,\tau_i}$ consistently estimate the population mean and covariance of the target, which can be implied by $0\in\mathcal{D}$.

\begin{assumption}\label{ass:a4}For $m=0,1,\dots,M$ and $i=1,\dots,[\widetilde{N}/h]$, the estimated covariance matrix at time $\tau_i$, denoted as $\widehat{\bm{\Sigma}}_{m,\tau_i}$, is positive definite almost surely.
\end{assumption}
In \hyperref[ass:a4]{Assumption 4}, we impose restrictions on covariance estimators at time $\tau_i$ for each dataset. Our methodological framework is based on portfolio Sharpe ratio optimization, which fundamentally requires strictly positive variance for all admissible investment strategies. We impose the assumption of an almost sure positive definiteness on the covariance estimator to satisfy the fact that for any column vector $\bm{x}\ne \bm{0}$, there exists $\bm{x}^{\top}\widehat{\bm{\Sigma}}_{m,\tau_i}\bm{x}>0$ almost surely.

\begin{theorem}\label{thm1}
   If \hyperref[ass:a1]{Assumptions 1}-\hyperref[ass:a4]{4} are satisfied, then
$$\Gamma(\bm{\widehat{w}})\rightarrow  1$$ in probability. 
\end{theorem}
This theorem demonstrates a kind of informative dataset selection consistency, in that our strategy can automatically assign all the weights to the datasets which can provide effective information for the target. It is worth to note that this theorem holds only for the case where the estimators calculated by the target data need to be helpful. This theorem also effectively avoids the problem of negative transfer. See Appendix B for the proof of \hyperref[thm1]{Theorem 1}.

Having established the convergence properties of the weight estimator, we now characterize the Sharpe ratio of the target portfolio at time $\widetilde{N}+1$ under our proposed strategy. Denote
$$\widetilde{SR}_{0,\tau_{i+1}}(\bm{w})= \frac{(\bm{\mu}_{0,\tau_{i+1}}^{\top}) \widehat{\bm{\phi}}_{\tau_{i+1}}(\bm{w})}{\sqrt{\widehat{\bm{\phi}}_{\tau_{i+1}}(\bm{w})^{\top}\bm{\Sigma}_{0,\tau_{i+1}}\widehat{\bm{\phi}}_{\tau_{i+1}}(\bm{w})}},\ \ \widetilde{MS}_{\widetilde{N}}(\bm{w})=\frac{1}{[\widetilde{N}/h]-1}\sum_{i=1}^{[\widetilde{N}/h]-1 }\widetilde{SR}_{0,\tau_{i+1}}(\bm{w}).$$ 

\begin{theorem}\label{thm2}If \hyperref[ass:a1]{Assumptions 1}-\hyperref[ass:a4]{4} are satisfied, then
$$\frac{\widetilde{SR}_{0,\widetilde{N}+1}(\widehat{\bm{w}})}{SR_{0,\widetilde{N}+1}}=1+o_p(1).$$
\end{theorem}
This theorem demonstrates that our strategy can
asymptotically obtain the maximum Sharpe ratio $SR_{0,\widetilde{N}+1}$. See Appendix C for the proof of \hyperref[thm2]{Theorem 2}. 

We have already examined the asymptotic behaviors of the weight estimators $\widehat{\bm{w}}$ and the resulting Sharpe ratio $\widetilde{SR}_{0,\widetilde{N}+1}(\widehat{\bm{w}})$. Considering that the traditional $Non-transfer$ strategy also asymptotically obtains the maximum Sharpe ratio $SR_{0,\widetilde{N}+1}$ under \hyperref[ass:a1]{Assumption 1} and \hyperref[ass:a3]{Assumption 3} (see Appendix C), we compare the variances between the Sharpe ratios of our proposed strategy and the $Non-transfer$ strategy. We further need the following assumptions.

\begin{assumption}\label{ass:a5} For $m=1,\dots,M$, $\underset{}{\text{limsup\ }}(N_0/N_m)\leq 1$.\end{assumption}
\hyperref[ass:a5]{Assumption 5} imposes certain restrictions on the sample size of each dataset. It is common and reasonable in practice, since the sample size of the target data is usually smaller. Given that all datasets in our analysis share the same sampling frequency, this assumption is very easy to be satisfied in our framework.

\begin{assumption}\label{ass:a6} There exists $\bm{w}^*\in\mathcal{W}$ such that $\widehat{\bm{w}}\rightarrow \bm{w}^*$ in probability.\end{assumption}
This assumption necessitates the convergence of the weight estimator, which can usually be achieved through an appropriate initial value and a fixed iterative direction. 

Let $\bm{\phi}^{[-d]}\in \text{R}^{d-1}$ denote the truncated vector comprising the first $d-1$ components of $\bm{\phi}\in \text{R}^d$.

\begin{assumption}\label{ass:a7} 
For $m,n\in\mathcal{D}$, $\parallel[N_m\text{Var}(\widehat{\bm{\phi}}^{[-d]}_{m,\widetilde{N}+1})][N_n\text{Var}(\widehat{\bm{\phi}}^{[-d]}_{n,\widetilde{N}+1})]^{-1}-\bm{I}_{d-1}\parallel_{\text{F}}\rightarrow 0$.
\end{assumption}
This assumption reveals the fundamental relationship between the variance of the estimator and the size of the training samples. Specifically, for $m,n\in\mathcal{D}$, the relationship between $\text{Var}(\widehat{\bm{\phi}}_{m,\tau_i})$ and $\text{Var}(\widehat{\bm{\phi}}_{n,\tau_i})$ is asymptotically determined by the sample sizes used for estimation. Referring to the Theorem $1$ in \citet{central}, we provide a proof of this assumption based on some regularity conditions in Appendix F. A concrete example can also be found in \hyperref[sec:3.3]{Simulation 1}, where the estimators derived from $S_1$ and $S_5$ exhibit precisely this property.

Denote
$$\widetilde{SR}_{0,\tau_{i+1}}= \frac{(\bm{\mu}_{0,\tau_{i+1}}^{\top}) \widehat{\bm{\phi}}_{0,\tau_{i+1}}}{\sqrt{\widehat{\bm{\phi}}_{0,\tau_{i+1}}^{\top}\bm{\Sigma}_{0,\tau_{i+1}}\widehat{\bm{\phi}}_{0,\tau_{i+1}}}},\ \ \widetilde{MS}_{\widetilde{N}}=\frac{1}{[\widetilde{N}/h]-1}\sum_{i=1}^{[\widetilde{N}/h]-1 }\widetilde{SR}_{0,\tau_{i+1}}
.$$

\begin{theorem}\label{thm3} If \hyperref[ass:a1]{Assumptions 1}-\hyperref[ass:a7]{7} are satisfied, then
$$\text{Var}(\widetilde{SR}_{0,\widetilde{N}+1}(\widehat{\bm{w}}))\leq\text{Var}(\widetilde{SR}_{0,\widetilde{N}+1})(1+o(1)).$$\end{theorem}
\hyperref[thm3]{Theorem 3} demonstrates the comparison of the volatility of the Sharpe ratio obtained using the $TL$ strategy and that of the $Non-transfer$ baseline strategy.  See Appendix D for the proof of \hyperref[thm3]{Theorem 3}.

\section{Simulation studies}\label{sec:3}
In this section, we conduct two simulation studies to evaluate the performance of our proposed investment strategy and empirically validate the theoretical results established in \hyperref[sec:2]{Section 2}. To thoroughly compare investment strategies, we employed two different data generation processes (DGP) in \hyperref[sec:3.3]{Simulation 1} and \hyperref[sec:3.4]{Simulation 2}, respectively. In \hyperref[sec:3.3]{Simulation 1}, we implement a standard experimental setup. First, we systematically compare the investment performance of the proposed $TL$ strategy with several benchmark strategies under progressively increasing distributional shifts between the target and some source datasets. Second, by incrementally expanding the sample size of each dataset, we empirically validate the theoretical properties of the $TL$ strategy, as derived in \hyperref[sec:2]{Section 2}. Then, we conduct a comparative analysis between our strategy and another transfer learning-based investment strategy proposed in \cite{cao2023ssrn} to assess their performance. Furthermore, based on empirical Fama French three factor model parameters derived from real market data, we simulate each dataset in \hyperref[sec:3.4]{Simulation 2} to conduct a rigorous comparative analysis across different investment strategies.

\subsection{Alternative portfolio strategies}\label{sec:3.1}
To assess the efficacy of our proposed strategy, we conduct a comprehensive comparative analysis with the following four established strategies in our simulation studies.

$TL_{equal}$: The target data and the source data are combined using equal weighting proportions in the transfer process. That is,
$$\widehat{\bm{\phi}}_{\widetilde{N}+1}=\sum\limits_{m=0}^M\frac{1}{M+1}\widehat{\bm{\phi}}_{m,\widetilde{N}+1}.$$

$Non-transfer$: Maximize the Sharpe ratio of the target assets. That is,
$$\widehat{\bm{\phi}}_{\widetilde{N}+1}=\underset{\bm{\phi} \in \Phi}{\arg \max } \frac{\widehat{\bm{\mu}}_{0,\widetilde{N}+1}^{\top} \bm{\phi}}{\sqrt{\bm{\phi}^{\top} \widehat{\bm{\Sigma}}_{0,\widetilde{N}+1} \bm{\phi}}}. $$

$Pool$: Combine the target and source datasets into a unified sample population and estimate the mean and covariance using the historical sample mean and historical sample covariance of the combined data. In detail, the estimated mean vector is
$$\widehat{\bm{\mu}}^*_{0,\widetilde{N}+1}=\frac{1}{N_0+\dots+ N_M}\sum\limits_{m=0}^M\sum\limits_{j=-(N_m-\widetilde{N}-1)}\limits^{\widetilde{N}} \bm{r}_{m,j},$$ and 
the estimated covariance matrix is $$\widehat{\bm{\Sigma}}^*_{0,\widetilde{N}+1}=\frac{1}{N_0+\dots+ N_M-1}\sum\limits_{m=0}^M\sum\limits_{j=-(N_m-\widetilde{N}-1)}\limits^{\widetilde{N}} (\bm{r}_{m,j}-\widehat{\bm{\mu}}^*_{0,\widetilde{N}+1})(\bm{r}_{m,j}-\widehat{\bm{\mu}}^*_{0,\widetilde{N}+1})^{\top}.$$Hence, the asset allocation can be estimated by $$\widehat{\bm{\phi}}_{\widetilde{N}+1}=\underset{\bm{\phi} \in \Phi}{\arg \max } \frac{\widehat{\bm{\mu}}_{0,\widetilde{N}+1}^{*\top} \bm{\phi}}{\sqrt{\bm{\phi}^{\top} \widehat{\bm{\Sigma}}^*_{0,\widetilde{N}+1} \bm{\phi}}}.$$

\subsection{Evaluation methodology}\label{sec:3.2}

We employ the following indicator to evaluate the out-of-sample forecasting performance across different strategies. Denote $\mathcal{O}$ as the set of time points designated for out-of-sample prediction.

$SSR$: Calculate the sample Sharpe ratio (SSR) of the portfolio strategy, defined as
$$SSR=\frac{\bar{E}(\widehat{\bm{\phi}}_t^{\top}\bm{r}_t)}{\sqrt{\bar{V}(\widehat{\bm{\phi}}_t^{\top}\bm{r}_t)}},\ t\in\mathcal{O},$$
where $\bar{E}(.)$ and $\bar{V}(.)$ represent the sample mean and sample variance, respectively.

\subsection{Simulation 1: Benchmark DGP}\label{sec:3.3}

Consider a capital market consisting of $5$ assets, that is, $d=5$. The DGP follows the simulation settings in \cite{r32}.
\begin{equation*}\begin{aligned}
\bm{r}_{m,t}=\bm{\alpha}+\bm{X}_{t}\bm{\Pi}_{m,t}+\bm{e}_{m,t},\ \ 
\bm{X}_{t+1}=\bm{BX}_{t}+\bm{\gamma}_{t+1},
\end{aligned}\tag{10}\label{10}\end{equation*}
where $\bm{r}_{m,t}\in \text{R}^5,\ \bm{X}_t\in \text{R}^{5\times3},\ \bm{\Pi}_{m,t}\in \text{R}^3,\ \bm{e}_{m,t}\in \text{R}^5,\ \bm{B}\in \text{R}^{5\times5},\ \bm{\gamma}_{t}\in \text{R}^{5\times3}$ and $\ m=0,\dots,5,\ t=1,\dots,N_m$. The term $\bm{\alpha}=(0.5,0.5,0.5,0.5,0.5)^{\top}$.
For the isolation term, $\bm{e}_{m,t}$ and every column of $\bm{\gamma}_t$ are independently identically distributed and follow a multivariate normal distribution $\text{MVN}(\bm{0},\bm{\Omega})$, where the components of the covariance matrix $\bm{\Omega}$ are $\bm{\Omega}(i,j)=0.5^{|i-j|}$. To ensure the stationarity of $\bm{X}_t$, the elements of the parameter matrix $\bm{B}$ are randomly generated from a uniform distribution over $(-1,1)$, with the additional constraint that all eigenvalues must lie within the unit circle. Regarding the regression parameters in this simulation, the following settings are used.
\begin{equation*}\begin{aligned}
&\bm{\Pi}_{0,t}=(0.9,0.6,0.7)^{\top},\ \ t=1,\dots,N_0,\\&\bm{\Pi}_{m,t}=(0.9,0.6,0.7)^{\top}+\frac{1}{t}\times\bm{\delta}_m,\ \ m=1,5,\ t=1,\dots,N_m,\\&\bm{\Pi}_{m,t}=(0.9,0.6,0.7)^{\top}+\rho\times\bm{\delta}_m,\ \ m=2,3,4,\ t=1,\dots,N_m,
\end{aligned}\end{equation*}where $\bm{\delta}_m \sim \text{MVN}(\bm{0},0.1\bm{I})$. We assign time-dependent parameters to the source data $S_1$ and $S_5$ in our simulation framework. On the one hand, it ensures that the historical sample mean and historical covariance matrix estimators for $S_1$ and $S_5$ satisfy \hyperref[ass:a1]{Assumption 1}. On the other hand, it allows these source parameters to asymptotically converge to their target counterparts, thereby ensuring that $S_1$ and $S_5$ provide effective transferable information. For $S_2$, $S_3$ and $S_4$, we introduce the term $\rho\times\bm{\delta}_m$ to the regression parameters $\bm{\Pi}_{m,t}$ to modulate the divergence between the source datasets and the target dataset.

\subsubsection{Investment performance under varying divergence}\label{3.3.1}
Recall that the scaling parameters $\{\rho\times\bm{\delta}_m|m=2,3,4\}$ govern the divergence between the target data and the non-informative source data. To evaluate the performance of each strategy, we systematically vary $\rho$ while holding ${\bm{\delta}_m}$ constant and compare the performance of each investment strategy in the part.

\begin{figure}[h] 
    \centering \includegraphics[width=320pt]{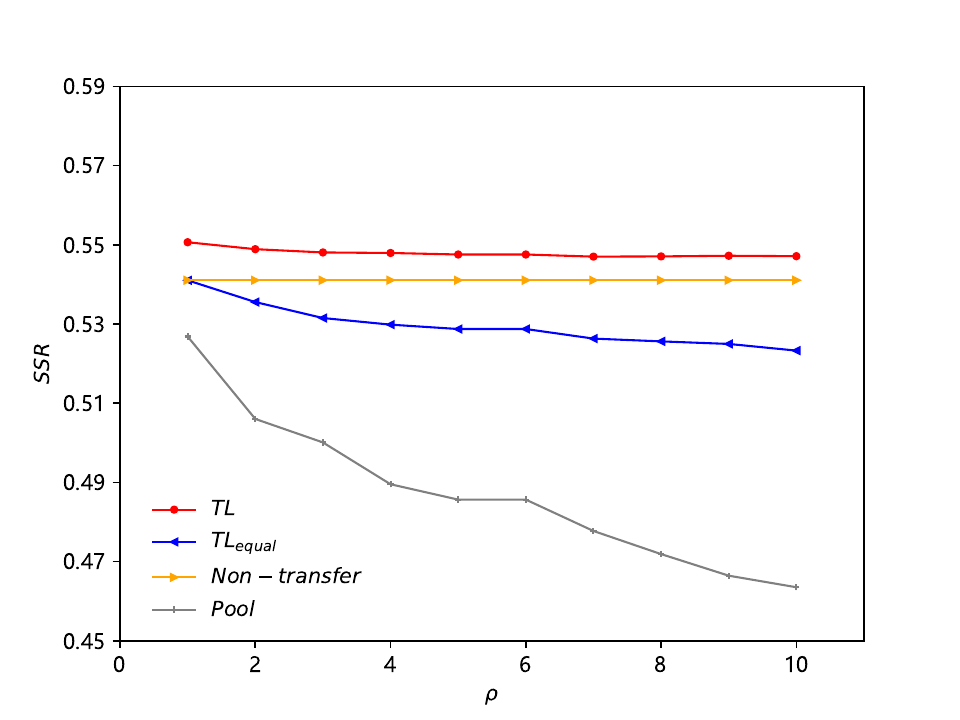}
    \label{SSRD}
    \captionsetup{singlelinecheck=off, justification=raggedright}
    \caption{The investment effect of each portfolio strategy}
  \raggedright\footnotesize
  This figure presents a comparative analysis of $SSR$ across $TL$ strategy (red dashed line with circular markers), $TL_{equal}$ strategy (blue solid line with left-pointing triangular markers), $Non-transfer$ strategy (orange solid line with right-pointing triangular markers) and $Pool$ strategy (gray solid line with star markers) as the divergence parameter $\rho$ increases from $1$ to $10$. The parameter $\rho$ modulates the distributional discrepancy between the non-informative source datasets $S_2$, $S_3$, $S_4$ and the target data $T$ systematically.
    \label{SSRbasic}
\end{figure}

Generate $N_0=500,\ N_m=(m+1)N_0$ samples for each dataset and use the final $50$ samples to do the out-of-sample forecasting. Repeat the total process $100$ times. Set $h$ to be $N_0/5$. As we incrementally increase $\rho$ to $10$, the $SSR$ values computed using each strategy are presented in \hyperref[SSRbasic]{Figure 3}.

\begin{figure}[h] 
    \centering     \includegraphics[width=320pt]{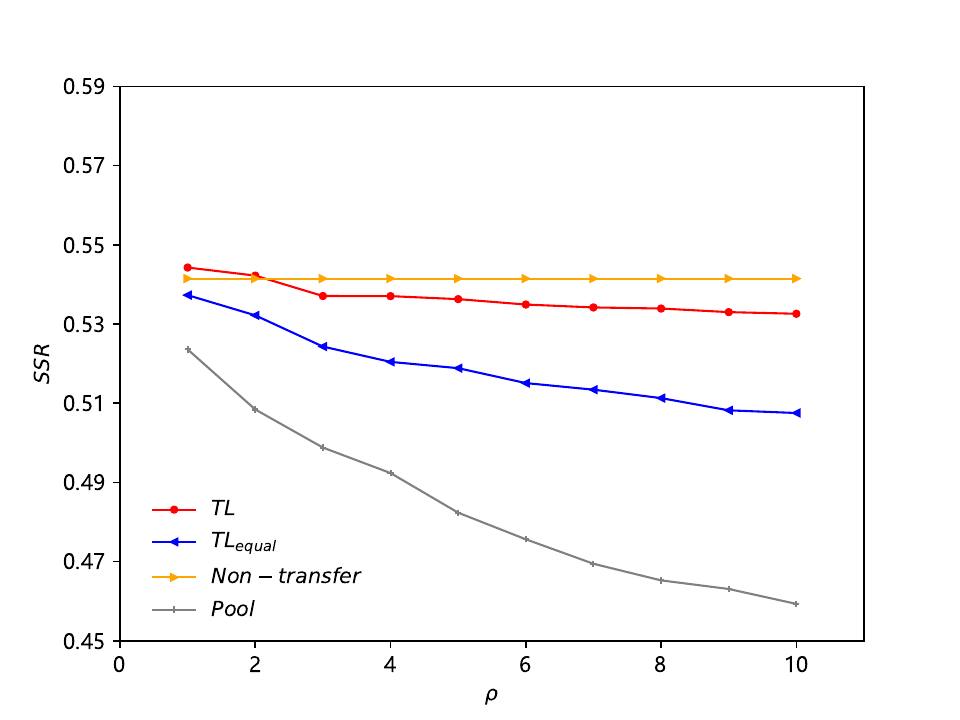}
        \label{SSRD_non}
        \captionsetup{singlelinecheck=off, justification=raggedright}
    \caption{The investment effect of each portfolio strategy while only utilizing the non-informative datasets $S_2$, $S_3$ and $S_4$}
  \raggedright\footnotesize
  This figure presents the $SSR$ across $TL$ strategy (red dashed line with circular markers), $TL_{equal}$ strategy (blue solid line with left-pointing triangular markers), $Non-transfer$ strategy (orange solid line with right-pointing triangular markers) and $Pool$ strategy (gray solid line with star markers) as the divergence parameter $\rho$ increases from $1$ to $10$ while utilizing the non-informative datasets $S_2$, $S_3$ and $S_4$ when $N_0=500$. 
\end{figure}

As evidenced by \hyperref[SSRbasic]{Figure 3}, the $TL$ strategy consistently achieves the maximal $SSR$ values among these investment strategies. Notably, the performance of the $Pool$ strategy deteriorates progressively as $\rho$ increases, suggesting that indiscriminate data merging becomes increasingly suboptimal when the divergence between the source and target data increases. Both the $TL$ strategy and the $TL_{equal}$ strategy exhibit a decreasing efficacy with increasing $\rho$, although $TL$ strategy maintains superior performance throughout. This shows that the targeted utilization of source data information, rather than naive aggregation, is crucial to optimize the investment decisions of the target assets and enhance the returns. Furthermore, the comparison of the $TL$ and $Non-transfer$ strategies reveals that the performance gap in $SSR$ decreases as $\rho$ increases, but the gap between the two strategies never vanishes. Accordingly, we investigate the investment performance for each strategy using only the non-informative source datasets $S_2$, $S_3$ and $S_4$ when $N_0=500$. As evidenced by \hyperref[SSRD_non]{Figure 4}, when utilizing only the non-informative datasets, the $TL$ strategy slightly outperforms the $Non-transfer$ strategy at $\rho = 1$. However, its performance deteriorates as $\rho$ increases. Once $\rho$ exceeds $2$, the $Non-transfer$ strategy emerges as superior, rendering the $TL$ strategy suboptimal under these conditions.

\subsubsection{Validity of the convergence of weight}\label{3.3.2}
To evaluate the capability to autonomously identify the datasets which can provide effective information for the target of the proposed $TL$ strategy, we calculate $\Gamma(\widehat{\bm{w}})$ at $N_0+1$ across varying sample sizes in this part.

Under different $N_0$, set $h$ to be $N_0/5$, $N_m=(m+1)N_0$ and $\rho$ to be $3$. Let $N_0$ be in the set $\{300,500,800,1000,1200,1500,2000,4000,8000,16000,40000\}$. It can be seen that $\mathcal{D}=\{0,1,5\}$, so $\Gamma(\widehat{\bm{w}})=\widehat{w}_0+\widehat{w}_1+\widehat{w}_5$. Using the specified experimental configuration, we generate synthetic datasets and evaluate $\Gamma(\widehat{\bm{w}})$ for target data at time $N_0+1$. Repeat the whole procedure $1000$ times.
\begin{figure}[h]
    \centering
\includegraphics[width=300pt]{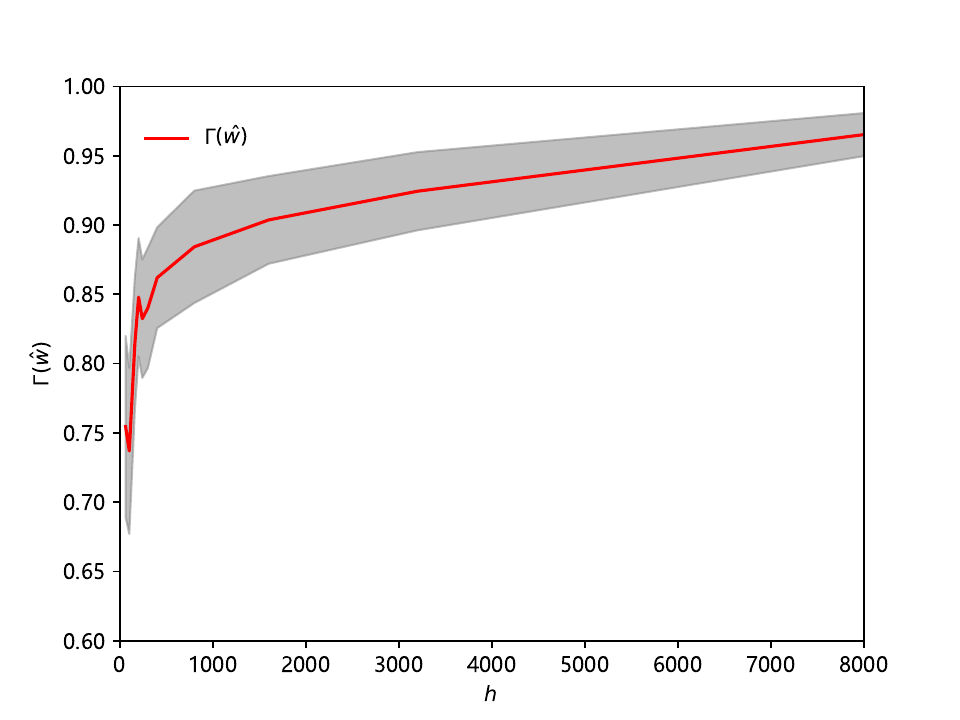}
    \captionsetup{singlelinecheck=off, justification=raggedright}
    \caption{$\Gamma(\widehat{\bm{w}})$ under different $h$}
    \raggedright
    \footnotesize This figure demonstrates the weight summation assigned to the datasets that can provide effective information under different estimation window $h$. The gray area is calculated by adding or subtracting $1.96$ standard error from the average value of $\Gamma(\widehat{\bm{w}})$ in $1000$ repetitions, which exhibits the corresponding variation of $\Gamma(\widehat{\bm{w}})$ in different $h$.
    \label{gamma_w}
\end{figure}

\hyperref[gamma_w]{Figure 5} shows $\Gamma(\widehat{\bm{w}})$ across different $h$. As shown, the aggregated weights assigned to informative datasets monotonically increase with sample size, asymptotically approaching $1$. This empirical result strongly corroborates the theoretical convergence established in \hyperref[thm1]{Theorem 1}.  
\begin{figure}[h] 
    \centering 
    
    \includegraphics[width=280pt]{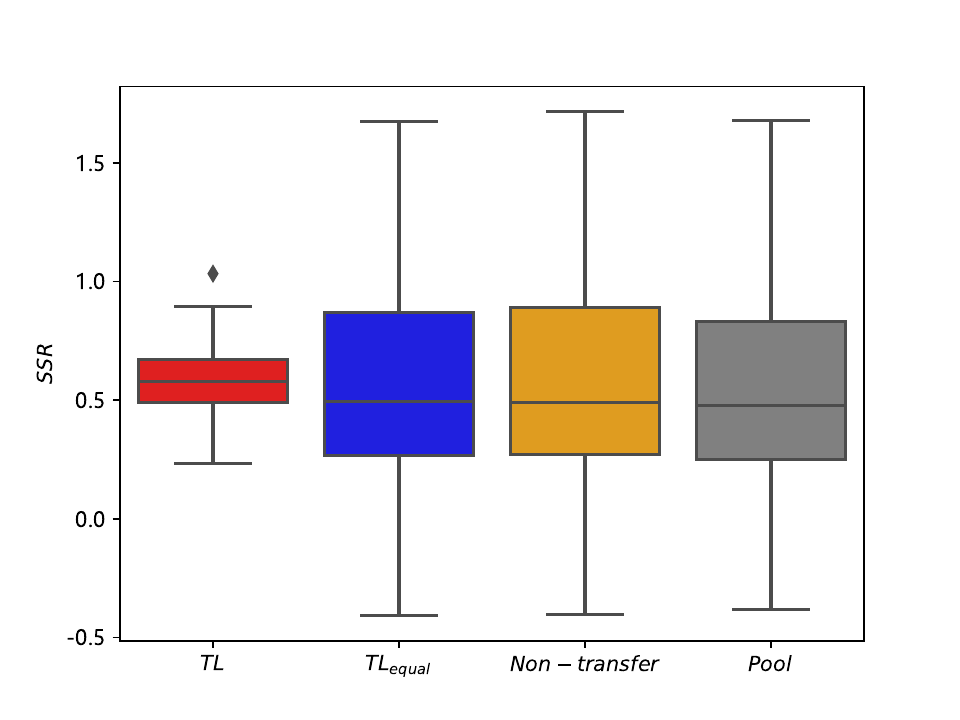}
\captionsetup{singlelinecheck=off, justification=raggedright}
    \caption{The box plots of $SSR$ obtained by various strategies}
    \raggedright
    \footnotesize This figure demonstrates box plots comparing $SSR$ across $TL$ strategy (red box plot), $TL_{equal}$ strategy (blue box plot), $Non-transfer$ strategy (orange box plot) and $Pool$ strategy (gray box plot). The horizontal lines within each box plot represent the sample mean of $SSR$ computed over $100$ simulation replicates, while the box widths correspond to the inter-quartile ranges, illustrating the variance in $SSR$ performance for each strategy.
    \label{5000ssr}
\end{figure}

\subsubsection{Comparison of the variance}

We further evaluate the performance of each strategy by computing $SSR$. For $N_0=5000$, use $h=N_0/5$, $N_m=(m+1)N_0$, respectively. The accuracy of the out-of-sample forecasting is evaluated using the final $1000$ target observations. From \hyperref[5000ssr]{Figure 6}, a smaller variance can be observed in the index $SSR$ when using the $TL$ strategy relative to other strategies, which can verify the conclusion of \hyperref[thm3]{Theorem 3} in this article. Under \hyperref[ass:a1]{Assumption 1}, we can show that the index $SSR$ converges almost surely to $\widetilde{SR}_{0,N_0+1}(\widehat{\bm{w}})$ as $N_0$ and the cardinality of $\mathcal{O}$ approaches infinity. Furthermore, leveraging the stationarity property of the target dataset, we can explicitly compute its population mean and covariance parameters. This enables the determination of the theoretically optimal Sharpe ratio, which is found to be $0.638$ for this target domain.

\subsubsection{Investment performance comparison: TL vs. TLc}\label{3.3.3}

To further evaluate the performance of our strategy, we conduct a comparative analysis with the portfolio optimization strategy based on transfer learning proposed in \cite{cao2023ssrn}. Their strategy incorporates the information from the source domain through a corrective framework, solving the following optimization problem to determine the optimal asset allocations.

$$\widehat{\bm{\phi}}_T=\underset{\bm{\phi} \in \Phi}{\arg \max } \frac{\widehat{\bm{\mu}}_T^{\top} \bm{\phi}}{\sqrt{\bm{\phi}^{\top} \widehat{\bm{\Sigma}}_T \bm{\phi}}}-\lambda\left\|\widehat{\bm{\phi}}_S-\bm{\phi}\right\|_2^2,$$ where $\widehat{\bm{\phi}}_T$ 
denotes the asset allocation of the target data and $\widehat{\bm{\phi}}_S$ denotes the asset allocation of the source data estimated using the $Non-transfer$ strategy. $\widehat{\bm{\mu}}_T$ and $\widehat{\bm{\Sigma}}_T$ are the historical sample mean and historical sample variance of the target data. Denote this strategy as $TLc$ strategy. Following \cite{cao2023ssrn}, we adopt the same regularization parameter $\lambda=0.2$ for comparative analysis. Given the constraint of single-source dataset utilization of the $TLc$ strategy, we evaluate the performance using both the target data and the previously generated single source dataset. Set $N_0$ to be $500$, $N_m=(m+1)N_0$ and $h$ to be $60$. Use the final $50$ samples to do the out-of-sample forecasting and repeat the total process $100$ times.

We first evaluate the strategies using $S_1$ as the source dataset, representing the case where the source data provides effective transferable information for the target domain. Since $S_1$ and $S_5$ follow identical distributional characteristics, the choice between them is substantively inconsequential. As demonstrated in \hyperref[SSR_TLc]{Table 1}, the $TL$ strategy is slightly better than the $TLc$ strategy.
\begin{table}[h]
    \centering
    \captionsetup{singlelinecheck=off, justification=raggedright}
    \caption{The investment effect of each portfolio strategy across different $N_0$}
    \begin{tabular}{cc}
\hline 
strategy& $SSR$\\
\hline $TL$ & \textit{0.550} (0.052)\\
$TL_{equal}$ &	0.549 (0.052)\\
$Non-transfer$ &0.541 (0.051)\\
$Pool$ & \textbf{0.551} (0.052)\\
$TL_c$&0.549 (0.052)\\
\hline
\end{tabular}
\\[5pt] 
\raggedright 
    \footnotesize{This table
    presents the $SSR$ performance comparing the proposed strategy with alternative strategies when utilizing the informative source dataset $S_1$. We bold the results of the best strategy and mark the results of the suboptimal strategy in italics.}
    \label{SSR_TLc}
\end{table} 

To further evaluate the performance of each strategy under non-informative source conditions, we introduce controlled variation in the source data and employ $S_3$ as the representative dataset. Notably, since $S_2$, $S_3$ and $S_4$ share identical distributional properties, the selection among them is substantively equivalent for our comparative analysis.
\begin{figure}[h] 
    \centering \includegraphics[width=320pt]{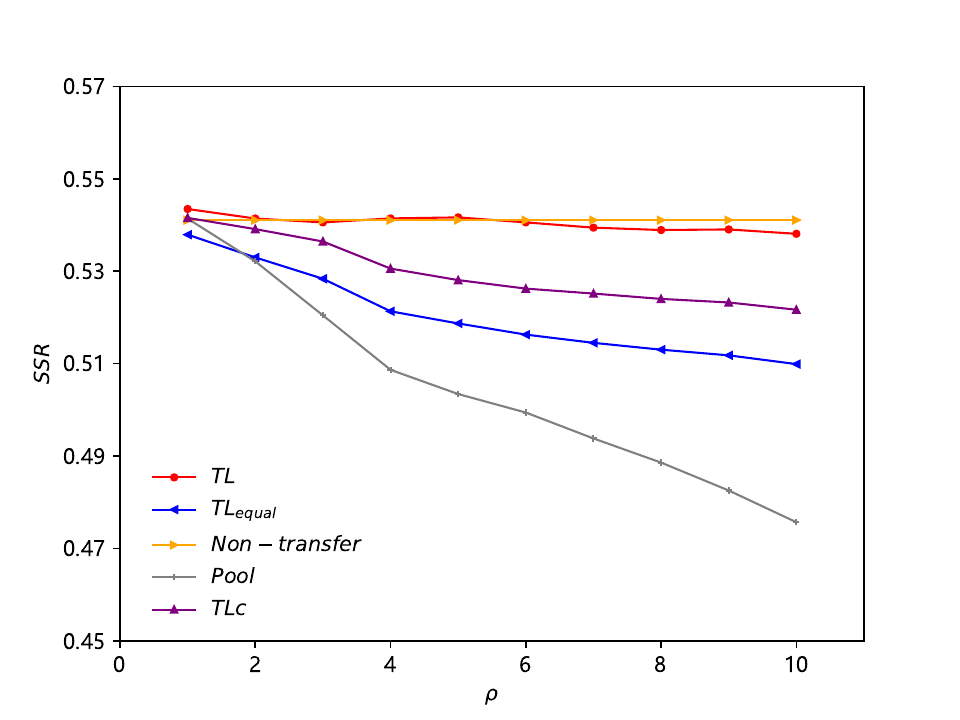}
        \label{cao_500_s3}
        \captionsetup{singlelinecheck=off, justification=raggedright}
    \caption{The investment effect of each portfolio strategy when only utilizing the non-informative dataset $S_3$}
  \raggedright\footnotesize
  This figure presents the $SSR$ performance comparing the proposed $TL$ strategy (red dashed line with circular markers) with alternative $TL_{equal}$ strategy (blue solid line with left-pointing triangular markers), $Non-transfer$ strategy (orange solid line with right-pointing triangular markers), $Pool$ strategy (gray solid line with star markers) and $TLc$ strategy (purple solid line with upward-pointing when utilizing source dataset $S_3$. 
    \label{cao_s3}
\end{figure}

\hyperref[cao_s3]{Figure 7} demonstrate the index $SSR$ of each strategy when the gap between $S_3$ and the target data increases when $N_0=500$. In general, the $TL$ strategy and 
$Non-transfer$ strategy perform better than others. When $\rho$ is larger, $TL$ strategy is slightly inferior than the $Non-transfer$ strategy. We can see that the proposed strategy consistently outperforms $TLc$ strategy across all $\rho$ values. Furthermore, the $TL$, $TL_{equal}$, $TLc$ and $Pool$ strategies exhibit a monotonically decreasing performance as $\rho$ increases.

\subsection{Simulation 2: FF3 model-based DGP}\label{sec:3.4}
To enhance the realism of our simulated data, we use empirical market data and estimate the Fama French three factor model ($FF3$) using the ordinary least squares strategy. The simulated data are then generated based on the estimated model parameters. The $FF3$ model, introduced by \citet{fama1993}, addresses the limitations of Capital Asset Pricing Model. Their seminal work demonstrates that the market beta alone cannot fully explain cross-sectional stock return variations while three firm-specific characteristics, denoted as market capitalization, book-to-market ratio, and earnings-to-price ratio, significantly improve explanatory power. The $FF3$ model is formally specified as$$r_{t}-r_f=\alpha+\beta_1 SMB_t+\beta_2 HML_t+\beta_3 MKT_t+\epsilon_{t},$$ where $r_{t}$ refers to the return of individual stocks and $r_f$ refers to the risk-free interest rate. $SMB$ represents the average return of the stock portfolio of small-market companies minus the average return of the stock portfolio of large-market companies (according to small-market effect). $HML$ refers to the portfolio return rate obtained by shorting companies with a high book value ratio and $MKT$ refers to the excess return of the market portfolio.
The three factors required ($MKT$, $SMB$, $HML$) are obtained from the $CRSP$ database for the period from January $2021$ to December $2023$. For our empirical analysis, we use the daily returns of the five American real estate industries, denoted as $EXPI$, $VICI$, $NMRK$, $INVH$, and $JLL$, and analyze their daily returns over the same $3$ year period from January $2021$ to December $2023$ ($753$ trading days). Each observation is indexed chronologically as $t=1,\dots,753$, where $t = 1$ corresponds to January $4$, $2021$ (the first trading day of $2021$). Given the daily return of the five stocks and the three factors, fit the $FF3$ model and we can get the parameters $\bm{\alpha}=(\alpha_1,\dots,\alpha_5)^{\top}\in \text{R}^5$ and $\{\bm{\beta}_p=(\beta_{1,p},\beta_{2,p},\beta_{3,p})^{\top}\in \text{R}^{3}|\  p=1,\dots,5\}$, respectively.  

We generate synthetic target and source data through the following DGP. $$
\bm{r}_{m,t}=\bm{\alpha}+\bm{X}\bm{\Pi}_{m,t}+\bm{e}_{m,t},\ m=0,1,\dots,5,\ t=1,\dots,N_m,$$
where $\bm{r}_{m,t}\in \text{R}^5,\ \bm{e}_{m,t}\in \text{R}^5\sim \text{MVN}(\bm{0},\bm{\Omega})$ and the components of the covariance matrix are $\bm{\Omega}(i,j)=0.5^{|i-j|}$.
For the definition of $\bm{X}$ and $\bm{\Pi}_{m,t}$, use the following settings.
$$\bm{X}=(\bm{\beta}_1,\bm{\beta}_2,\bm{\beta}_3,\bm{\beta}_4,\bm{\beta}_5)^{\top}\in \text{R}^{5\times 3},$$
$$\bm{\Pi}_{m,t}=(SMB_t,HML_t,MKT_t)^{\top},\ m=0,\ t=1,\dots,N_0,$$
$$\bm{\Pi}_{m,t}=(SMB_t,HML_t,MKT_t)^{\top}+\frac{1}{t}\times\bm{\epsilon}_{m,t},\ m=1,5,\ t=1,\dots,N_m,$$
$$\bm{\Pi}_{m,t}=(SMB_t,HML_t,MKT_t)^{\top}+\rho \times\bm{\epsilon}_{m,t},\ m=2,3,4,\ t=1,\dots,N_m,$$
where $\bm{\epsilon}_{m,t}\in \text{R}^3\sim \text{MVN}(\bm{0},0.1\bm{I})$.

Generate $N_0=\dots=N_5=500$ samples for each dataset. Take $h=N_0/5$ and use the final $50$ samples to do the out-of-sample forecasting. Repeat the total process $100$ times.\footnote{To ensure temporal consistency across all datasets, we maintain identical sample sizes for both the factor data and stock returns in our simulation.}
\begin{figure}[h] 
    \centering 
    \includegraphics[width=320pt]{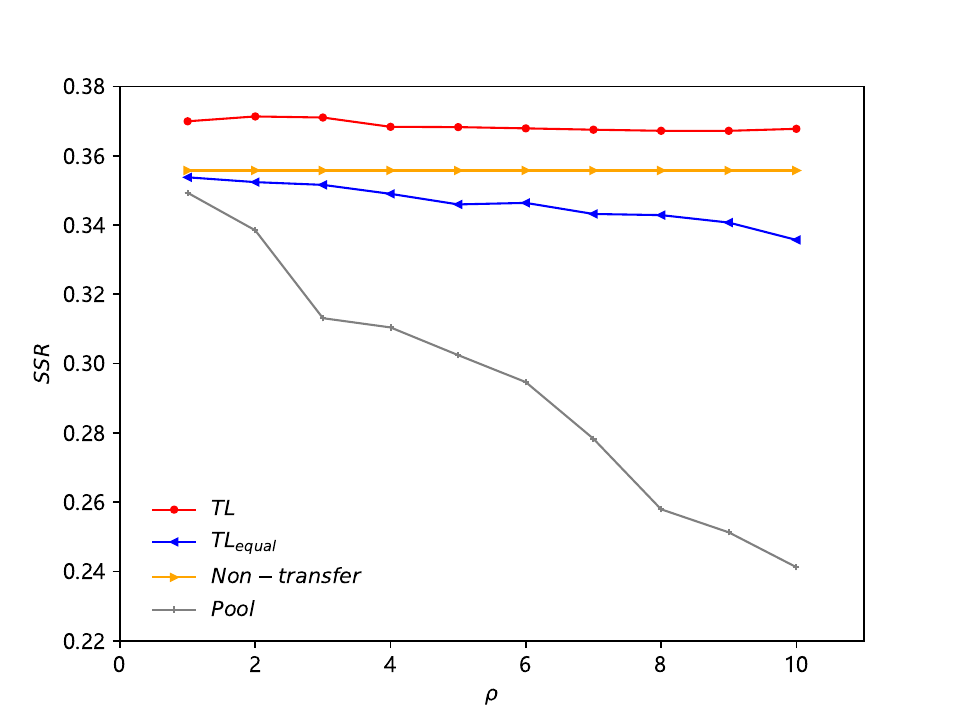}
        \label{FF3_500}
    \captionsetup{singlelinecheck=off, justification=raggedright}
    \caption{The investment effect of each portfolio strategy}
  \raggedright\footnotesize
  This figure presents a comparative analysis of $SSR$ across $TL$ strategy (red dashed line with circular markers), $TL_{equal}$ strategy (blue solid line with left-pointing triangular markers), $Non-transfer$ strategy (orange solid line with right-pointing triangular markers) and $Pool$ strategy (gray solid line with star markers) as the divergence parameter $\rho$ increases from $1$ to $10$. The parameters for the DGP of each dataset are determined by fitting the $FF3$ model. The parameter $\rho$ systematically modulates the distributional discrepancy between the non-informative source datasets $S_2$, $S_3$, $S_4$ and the target data $T$.
    \label{FF3}
\end{figure}

\hyperref[FF3]{Figure 8} presents $SSR$ performance across all evaluated strategies, using parameters calibrated from empirical market data. These findings are consistent with the results presented in \hyperref[SSRbasic]{Figure 3}. As shown, our approach demonstrates a clear advantage. Furthermore, the data generated for each dataset in this simulation do not necessarily satisfy \hyperref[ass:a1]{Assumption 1}. Despite this relaxation, the proposed $TL$ strategy still has obvious advantages, which also reflects the robustness of our strategy.

\section{Applications}\label{sec:4}

We evaluate the performance of our strategy using two distinct datasets: one comprising dual-listed A-share and H-share stocks, and the other consisting of stocks from various industrial sectors of the United States. The results based on the two stock universes are reported in \hyperref[sec:4.1]{Section 4.1} and \hyperref[sec:4.2]{Section 4.2}, respectively.

\subsection{Optimizing portfolios across H-shares and A-shares}\label{sec:4.1}

Many leading Chinese enterprises are dually listed on both the Shanghai/Shenzhen Stock Exchanges (A-shares) and the Hong Kong Stock Exchange (H-shares). This arrangement offers a valuable natural laboratory for our research. Although A-shares and H-shares represent claims on the same underlying firms, differences in investor composition, market regulations, and capital controls cause them to exhibit distinct price discovery dynamics and risk appetite. Rather than being mere noise, these differences create a solid foundation for informational complementarity.

From the perspective of using Hong Kong market data to inform A-share analysis, the highly internationalized and institution-dominated Hong Kong market tends to respond more swiftly and accurately to global macroeconomic trends, geopolitical developments, and sector-specific shifts. As a result, H-share price movements often incorporate global information that has not yet been fully reflected in the more retail-driven and relatively insulated A-share market. Incorporating H-share returns as a leading indicator into predictive models acts as an effective information set, capturing globally sourced fundamental changes and thereby improving the foresight and accuracy of A-share return forecasts. For example, in the cases of commodity firms or large technology companies, H-share prices may earlier reflect shifts in global supply-demand conditions or technological disruptions, thereby aiding in the more precise estimation of subsequent A-share performance.

Conversely, using A-shares data to interpret H-shares dynamics is equally valuable. The A-share market, with its substantial retail investor presence, serves as a direct barometer of domestic investor sentiment and reactions to local policy changes. Subtle shifts relating to local consumption trends, industrial policies, or regional risks often emerge first in A-share prices. For Mainland companies listed in Hong Kong, these locally-driven sentiments and expectations represent critical dimensions that international investors may overlook or respond to with a lag. Thus, A-share return dynamics provide essential information for understanding H-shares behavior.

Through bidirectional modeling, researchers can better disentangle global versus local drivers of firm value. This approach facilitates a deeper analysis of the the asset returns. In summary, integrating data from both markets combines two distinct information sets to form a more comprehensive and robust view of the target asset. We collect earnings data for five listed companies from four sectors: Energy, Manufacturing, Finance, and Medical, across both H-shares and A-shares. See Appendix G for the detailed information of the listed companies. 
\begin{table}[h]
    \centering
    \captionsetup{singlelinecheck=off, justification=raggedright}
    \caption{The investment effect of each portfolio strategy in different target markets}
    \begin{tabular}{ccccc}
\hline \multirow{2}{*}{\diagbox{Strategy}{Target market}} & \multicolumn{2}{c}{Energy}  & \multicolumn{2}{c}{Manufacturing}\\
\cline { 2 - 5 } & A-shares& H-shares& A-shares& H-shares\\
\hline $TL$ & \textbf{0.129} & \textbf{0.448} &\textbf{0.138} &\textit{0.191}\\
$TL_{equal}$ &0.101 & 0.394&0.118 &0.149\\
$Non-transfer$ & \textit{0.128} & 0.364&\textit{0.124} &\textbf{0.196}\\
$Pool$ &0.111  &0.358 &0.069 &0.175\\
$TLc$ &0.085 &\textit{0.413} &0.093 &0.185\\
\hline \multirow{2}{*}{\diagbox{Strategy}{Target market}} & \multicolumn{2}{c}{Financial}  &  \multicolumn{2}{c}{Medical}\\
\cline { 2 - 5 } & A-shares& H-shares& A-shares& H-shares\\
\hline $TL$ &\textbf{0.132} &\textit{0.244} &\textbf{0.186} &\textbf{0.968} \\
$TL_{equal}$ &0.125 & 0.237&0.177 &0.728 \\
$Non-transfer$ &0.130 &0.219 &\textit{0.183}  &\textit{0.753} \\
$Pool$ &0.113 & \textbf{0.252}&0.154  &0.651 \\
$TLc$ &\textit{0.131} & 0.241&0.148 & 0.553\\
\hline
\end{tabular}
\\[5pt]
\begin{minipage}{\linewidth} 
\raggedright 
    \footnotesize{This table demonstrates the $SSR$ index gained using the proposed strategy and the alternative strategies when investing in five listed companies from four sectors: Energy, Manufacturing, Finance, and Medical, across both H-shares and A-shares. For the convenience of comparison, we bold the results of the best strategy and mark the results of the suboptimal strategy in italics.}
    \end{minipage}
    \label{A-H}
\end{table}

Our analysis alternately designates one market as the target and the other as the source market. Take their daily return from July 2021 to June 2025 for analysis and the data from January 2025 to June 2025 are used to make out-of-sample forecasts. In detail, every dataset covers $968$ time periods and $|\mathcal{O}|=117$. The data information comes from $Compustat$ database.  Building on these comprehensive datasets, we conduct a comparative performance analysis between our proposed strategy and established benchmark strategies for the construction of the target assets portfolio. \hyperref[A-H]{Table 2} presents the comparative performance of portfolio strategies across different industry sectors. Overall, the method proposed in this article has performed quite well when investing in the H-shares and A-shares of different industries. The complementary nature of the information between these two markets can generate substantial added value for investors. \hyperref[weight]{Table 3} demonstrates the average transferring weight of different target markets within the period of out-of-sample forecasting when the $TL$ strategy is adopted. As shown, the source dataset has been effectively utilized in the $TL$ strategy we proposed.

\begin{table}[h]
            \centering
            \captionsetup{singlelinecheck=off, justification=raggedright}
    \caption{The average transferring weight of different target markets when $TL$ strategy is adopted}
            \begin{tabular}{cccc}
            \hline
                 \multicolumn{2}{c}{\diagbox{Target market}{Weights}}&$w_0$&$w_1$\\
                 \hline
                 \multirow{2}{*}{Energy}&A-shares&0.970 &0.030\\ 
                 &H-shares&0.632 &0.368\\
                 \multirow{2}{*}{Manufacturing}&A-shares&0.854&0.146\\
                &H-shares&0.837 &0.163\\
                \multirow{2}{*}{Financial}&A-shares&0.751 &0.249\\
                &H-shares&0.872 &0.128 \\
                \multirow{2}{*}{Medical}&A-shares&0.852 &0.148 \\
                &H-shares&0.743  & 0.257\\ \hline   
            \end{tabular}
            \\[5pt]
\begin{minipage}{\linewidth} 
\raggedright 
    \footnotesize{This table demonstrates the average transferring weight of different target markets within the period of out-of-sample forecasting when the $TL$ strategy is adopted. As mentioned before, $w_0$ is the weight assigned in the target market and $w_1$ is the weight assigned in the source market.}
    \end{minipage}
    \label{weight}
        \end{table}

\subsection{Optimizing portfolios across sectors}\label{sec:4.2}

The real estate sector constitutes a fundamental pillar of modern economic systemsand plays a critical role in both developed and emerging economies. As a tangible assetclass, residential properties provide not only shelter but also represent one of the mostsigniffcant components of household wealth portfolios globally. The sector also offersmultiple pathways for value creation, including capital appreciation through strategicacquisitions and stable income generation via rental operations. As a linchpin sector with extensive backward and forward links, real estate exhibitsremarkable stimulating effects across upstream industries (e.g., construction materials,steel production, and heavy machinery) and downstream sectors (e.g., interior design,home furnishings, and appliance manufacturing). The financial phenomenon between different sectors became particularly evident during the global ffnancial crisis in 2008,where mortgage-backed securities and real estate derivatives ampliffed systemic riskthrough complex ffnancial channels. This intricate web of inter-sectoral dependenciesraises a critical research question: can stock return data from correlated sectors beleveraged to enhance the investment performance of the interested market?
\begin{table}[h]
\centering
\captionsetup{singlelinecheck=off, justification=raggedright}
\caption{The investment effect of each portfolio strategy in different target markets}
\begin{tabular}{cccc}
\hline \multirow{2}{*}{} & Real Estate & Financial & Construction \\
\hline $TL$ & \textit{0.129} & \textbf{0.354} & \textbf{0.089} \\
$TL_{equal}$ & \textbf{0.162} & \textit{0.199} & \textit{0.064} \\
$Non-transfer$ & -0.004 & 0.198 & -0.072 \\
$Pool$ & 0.015 & 0.161 & 0.020 \\
\hline \multirow{2}{*}{} & Furniture & Manufacturing & Marketing \\
\hline
$TL$ & \textbf{-0.002} & \textit{0.181} & \textbf{0.084} \\
$TL_{equal}$ & \textit{-0.018} & 0.129 & 0.029 \\
$Non-transfer$ & -0.046 & -0.012 & \textit{0.053} \\
$Pool$ & -0.090 & \textbf{0.303} & -0.001 \\
\hline
\end{tabular}
\\[5pt]
\begin{minipage}{\linewidth} 
\raggedright 
    \footnotesize{This table demonstrates the $SSR$ performance comparing our proposed strategy with alternative strategies when investing in the real estate industry, financial industry, construction industry, furniture industry, manufacturing industry and marketing industry in the American stock market. In each setting, we bold the results of the best strategy and mark the results of the suboptimal strategy in italics.}
    \end{minipage}
    \label{TL_industry}
\end{table}

We implement a comprehensive cross-sector analysis using the real estate industry
data, financial industry data, construction industry data, furniture industry data, manufacturing industry data and marketing industry data in American stock market, systematically rotating each industry sector as the target domain while employing all other sectors as source domains. Select the five largest market capitalization stocks from both target and the source industries. Take their daily return from January $2020$ to December $2023$ for analysis and the data from October $2023$ to December $2023$ are used to make out-of-sample forecasts. In detail, every dataset covers $877$ time periods and $|\mathcal{O}|=63$. The data information comes from $Osiris$, $CRSP$ and $Compustat$ database. Building on these comprehensive datasets, we conduct a comparative performance
analysis between our proposed strategy and established benchmark strategies for the construction of the target assets portfolio. 
\begin{table}[h]
\centering
\captionsetup{singlelinecheck=off, justification=raggedright}
\caption{The average transferring weight of each dataset when $TL$ strategy is adopted in different target markets}
\begin{tabular}{ccccccc}
\hline 
\diagbox{}{Target market}& Real Estate & Financial & Construction\\
\hline Real Estate & 0.523 & 0.236 & 0.061 \\
Financial & 0.175 & 0.451 & 0.018\\
Construction & 0.000 & 0.058 & 0.349\\
Furniture & 0.243 & 0.255 & 0.000\\
Manufacture & 0.000 & 0.000 & 0.519\\
Marketing & 0.059 & 0.000 & 0.053\\
\hline
\diagbox{}{Target market}& Furniture & Manufacture & Marketing \\
\hline Real Estate & 0.219 & 0.047 & 0.034 \\
Financial & 0.000 & 0.000 & 0.000 \\
Construction & 0.000 & 0.514 & 0.218 \\
Furniture & 0.781 & 0.008 & 0.082 \\
Manufacture & 0.000 & 0.431 & 0.638 \\
Marketing & 0.000 & 0.000 & 0.028 \\
\hline
\end{tabular}
\\[5pt]
\begin{minipage}{\linewidth} 
\raggedright 
\footnotesize{This table demonstrates the average transferring weight of different target markets within the period of out-of-sample forecasting when the $TL$ strategy is adopted. Each column of this table quantifies the transferring weight of each dataset (listed on the left) when a specific dataset serves as the target.}
    \end{minipage}
    \label{weight_industry}
\end{table}

\hyperref[TL_industry]{Table 4} presents the comparative performance
of portfolio strategies across different industry sectors. The TL strategy achieves either optimal or near-optimal $SSR$ in the tested industry pairs, while $Non-transfer$ strategy consistently underperforms. These findings provide robust empirical evidence that cross-industry data in American equity markets contain economically significant predictive signals for investment decision making. \hyperref[weight_industry]{Table 5} demonstrates the average transferring weight of different target markets within the period of out-of-sample forecasting when the $TL$ strategy is adopted. As shown, the source dataset in this example has also been effectively utilized in the $TL$ strategy we proposed.
\section{Conclusion}\label{sec:5}

In this article, we develop a novel transfer learning framework for portfolio optimization that systematically leverages cross-domain information to improve the investment performance of the target. Our theoretical analysis establishes that the proposed strategy possesses the weight consistency property and asymptotically achieves the maximum Sharpe ratio while maintaining a smaller variance than the conventional $Non-transfer$ strategy. The proposed strategy is relatively simple and easy to implement.

This study adopts the Sharpe ratio as our primary performance metric, which differs from the conventional mean-variance framework prevalent in the literature.  In the absence of estimation errors, the Sharpe ratio maximization and utility maximization are equivalent. But the Sharpe ratio is simpler because it does not require information about risk aversion parameters and is widely used by researchers and practitioners to compare trading strategies and models. In our implementation, we set the parameter $h=N_0/5$. Importantly, the theoretical guarantees of the proposed $TL$ strategy remain valid even when $N_0/h\rightarrow\infty$, as demonstrated in the Appendix. Although the choice of $h$ does not affect the theoretical properties of the strategy, determining its optimal selection remains an open question with considerable research significance, offering a promising direction for future investigations.

\clearpage
\setlength{\bibhang}{0pt}
\bibliography{sample}

@ARTICLE{r1,
author = {Markowitz, Harry M},
title = {PORTFOLIO SELECTION},
year = {1952},
journal = {Journal of Finance},
volume = {7},
number = {1},
pages = {77–91}
}

@ARTICLE{r3,
author = {Ledoit, Olivier and Wolf, Michael},
title = {A well-conditioned estimator for large-dimensional covariance matrices},
year = {2004},
journal = {Journal of Multivariate Analysis},
volume = {88},
number = {2},
pages = {365–411}
}

@ARTICLE{r4,
author = {Ledoit, Olivier and Wolf, Michael},
title = {Nonlinear shrinkage of the covariance matrix for portfolio selection: Markowitz meets goldilocks},
year = {2018},
journal = {Review of Financial Studies},
volume = {31},
number = {4},
pages = {1604-1604}
}

@article{r5,
author = {Bickel, Peter J. and Levina, Elizaveta},
title = {Covariance regularization by thresholding},
volume = {36},
journal = {Annals of Statistics},
number = {6},
publisher = {Institute of Mathematical Statistics},
pages = {2577-2604},
year = {2008}
}

@ARTICLE{r6,
author = {Fan, Jianqing and Fan, Yingying and Lv, Jinchi},
title = {High dimensional covariance matrix estimation using a factor model},
year = {2008},
journal = {Journal of Econometrics},
volume = {147},
number = {1},
pages = {186–197}
}

@article{r7,
author = { Fan, Jianqing and  Liu,Han and Wang, Weichen},
title = {Large covariance estimation through elliptical factor models},
volume = {46},
journal = {Annals of Statistics},
number = {4},
pages = {1383-1414},
year = {2015}
}

@ARTICLE{r10,
author = {Jagannathan, Ravi and Ma, Tongshu},
title = {Risk Reduction in Large Portfolios: Why Imposing the Wrong Constraints Helps},
year = {2003},
journal = {Journal of Finance},
volume = {58},
number = {4},
pages = {1651–1684}
}

@ARTICLE{r12,
author = {Fan, Jianqing and Zhang, Jingjin and Yu, Ke},
title = {Vast portfolio selection with gross-exposure constraints},
year = {2012},
journal = {Journal of the American Statistical Association},
volume = {107},
number = {498},
pages = {592–606}
}

@article{r16,
  title={Why Naive Diversification Is Not So Naive, and How to Beat It?},
  author={Yuan, Ming and Zhou, Guofu},
  journal={Journal of Financial and Quantitative Analysis},
  volume={59},
  number={8},
  pages={3601-3632},
  year={2024}
}

@ARTICLE{r17,
author = {W. F. Sharpe},
title = {The Sharpe ratio},
year = {1994},
journal = {Journal of Portfolio Management},
volume = {21},
pages = {49-58}
}

@article{r28,
Author = {Kraus, Mathias and Feuerriegel, Stefan},
Title = {Decision support from financial disclosures with deep neural networks
   and transfer learning},
Journal = {Decision Support Systems},
Year = {2017},
Volume = {104},
Pages = {38-48}
}

@article{ r29,
Author = {Li, Wei and Ding, Shuai and Chen, Yi and Wang, Hao and Yang, Shanlin},
Title = {Transfer learning-based default prediction model for consumer credit in
China},
Journal = {Journal of Supercomputing},
Year = {2019},
Volume = {75},
Number = {2, SI},
Pages = {862-884}
}

@article{r30,
title = {Jointly modeling transfer learning of industrial chain information and deep learning for stock prediction},
journal = {Expert Systems with Applications},
volume = {191},
pages = {116257},
year = {2022},
author = {Wu,Dingming  and  Wang,Xiaolong and Wu,Shaocong}
}

@article{r32,
author = {Filip, Klimenka and James, Lewis Wolter},
title = {Multiple Regression Model Averaging and the Focused Information Criterion With an Application to Portfolio Choice},
journal = {Journal of Business $\&$ Economic Statistics},
volume = {37},
number = {3},
pages = {506-516},
year = {2019}
}

@article{pan2010,
  author={Pan, Sinno Jialin and Yang, Qiang},
  journal={IEEE Transactions on Knowledge and Data Engineering}, 
  title={A Survey on Transfer Learning}, 
  year={2010},
  volume={22},
  number={10},
  pages={1345-1359},
}

@article{zhuang2011,
  author={Zhuang, Fuzhen and Qi, Zhiyuan and Duan, Keyu and Xi, Dongbo and Zhu, Yongchun and Zhu, Hengshu and Xiong, Hui and He, Qing},
  journal={Proceedings of the IEEE}, 
  title={A Comprehensive Survey on Transfer Learning}, 
  year={2021},
  volume={109},
  number={1},
  pages={43-76},
}

@article{jeong2019,
title = {Improving financial trading decisions using deep Q-learning: Predicting the number of shares, action strategies, and transfer learning},
journal = {Expert Systems with Applications},
volume = {117},
pages = {125-138},
year = {2019},
author = {Gyeeun Jeong and Ha Young Kim},
}

@article{morstedt2024,
title = {Cross validation based transfer learning for cross-sectional non-linear shrinkage: A data-driven approach in portfolio optimization},
journal = {European Journal of Operational Research},
volume = {318},
number = {2},
pages = {670-685},
year = {2024},
author = {Torsten Mörstedt and Bernhard Lutz and Dirk Neumann},
}

@article{hu2023,
  author  = {Hu,Xiaonan  and Zhang,Xinyu },
  title   = {Optimal Parameter-Transfer Learning by Semiparametric Model Averaging},
  journal = {Journal of Machine Learning Research},
  year    = {2023},
  volume  = {24},
  number  = {358},
  pages   = {1-53}
}

@article{koshiyama2022,
author = {Adriano Koshiyama and Stefano B. Blumberg and Nick Firoozye and Philip Treleaven and Sebastian Flennerhag},
title = {QuantNet: transferring learning across trading strategies},
journal = {Quantitative Finance},
volume = {22},
number = {6},
pages = {1071-1090},
year = {2022}
}

@article{zhang2024,
author = {Zhang,Xinyu  and Liu,Huihang  and Wei,Yizheng  and Ma,Yanyuan},
title = {Prediction Using Many Samples with Models Possibly Containing Partially Shared Parameters},
journal = {Journal of Business $\&$ Economic Statistics},
volume = {42},
number = {1},
pages = {187-196},
year = {2024}
}

@article{hansen2007,
author = {Hansen, Bruce E.},
title = {Least Squares Model Averaging},
journal = {Econometrica},
volume = {75},
number = {4},
pages = {1175-1189},
year = {2007}
}

@article{enrique2015,
  title={Model averaging in economics: An overview},
  author={Moral-Benito, Enrique},
  journal={Journal of Economic Surveys},
  volume={29},
  number={1},
  pages={46-75},
  year={2015}
}

@article{buccheri2021,
Author = {Buccheri, Giuseppe and Corsi, Fulvio and Peluso, Stefano},
Title = {High-Frequency Lead-Lag Effects and Cross-Asset Linkages: A Multi-Asset
   Lagged Adjustment Model},
Journal = {Journal of Business $\&$ Economic Statistics},
Year = {2021},
Volume = {39},
Number = {3},
Pages = {605-621},
}

@article{ledoit2017,
    author = {Ledoit, Olivier and Wolf, Michael},
    title = {Nonlinear Shrinkage of the Covariance Matrix for Portfolio Selection: Markowitz Meets Goldilocks},
    journal = {Review of Financial Studies},
    volume = {30},
    number = {12},
    pages = {4349-4388},
    year = {2017}
}

@article{ao2018,
    author = {Ao, Mengmeng and Li,Yingying and Zheng, Xinghua},
    title = {Approaching Mean-Variance Efficiency for Large Portfolios},
    journal = {Review of Financial Studies},
    volume = {32},
    number = {7},
    pages = {2890-2919},
    year = {2018}
}

@article{babi2023,
    author = {Bali, Turan G and Beckmeyer, Heiner and Mörke, Mathis and Weigert, Florian},
    title = {Option Return Predictability with Machine Learning and Big Data},
    journal = {Review of Financial Studies},
    volume = {36},
    number = {9},
    pages = {3548-3602},
    year = {2023}
}

@article{guo2025,
    author = {Guo, Hongye},
    title = {Earnings Extrapolation and Predictable Stock Market Returns},
    journal = {Review of Financial Studies},
    volume = {38},
    number = {6},
    pages = {1730-1782},
    year = {2025}
}

@article{sharpe1966,
    author = {William F. Sharpe},
    title = {Mutual Fund Performance},
    journal = {Journal of Business} ,
    year = {1966},
    volume = {39},
    number = {1},
    pages = {119–138}
}

@article{fama1993,
  title={Common risk factors in the returns on stocks and bonds},
  author={Fama, Eugene F and French, Kenneth R},
  journal={Journal of Financial Economics},
  volume={33},
  number={1},
  pages={3-56},
  year={1993}
}

@article{diebold2001,
  title={The distribution of realized stock return volatility},
  author={Diebold, F and Andersen, T and Bollerslev, T and Ebens, H},
  journal={Journal of Financial Economics},
  volume={61},
  pages={43-76},
  year={2001}
}

@book{cochrane2009,
  title={Asset pricing: Revised edition},
  author={Cochrane, John H},
  year={2009},
  publisher={Princeton university press}
}

@misc{lasse2024,
title={How global is predictability? \text{T}he power of financial transfer learning}, 
author={Hellum, Oliver and Pedersen, Lasse Heje and Ronn-Nielsen, Anders},
year={2024},
howpublished={doi:10.2139/ssrn.4620157. Woking Paper, SSRN, https://ssrn.com/abstract=4620157}
}

@article{kelly2023,
  title={Principal portfolios},
  author={Kelly, Bryan and Malamud, Semyon and Pedersen, Lasse Heje},
  journal={Journal of Finance},
  volume={78},
  number={1},
  pages={347-387},
  year={2023},
  publisher={Wiley Online Library}
}

@misc{cao2023ssrn,
	title={Risk of transfer learning and its applications in finance},
	author={Cao, Haoyang and Gu, Haotian and Guo, Xin and Rosenbaum, Mathieu},
	year={2023},
	howpublished={doi:10.2139/ssrn.4624427. Woking Paper, SSRN, https://ssrn.com/abstract=4624427}
}

@incollection{Zitikis2006,
author = {Zitikis, Ričardas and Zolatarev, Vladimir M. and Kalashnikov, Vladimir V.},
booktitle = {Stability Problems for Stochastic Models},
copyright = {Springer-Verlag 1993},
isbn = {3540567445},
issn = {0075-8434},
language = {eng},
publisher = {Springer Berlin Heidelberg},
title = {A berry-esséen bound for multivariate l-estimates with explicit dependence on dimension},
year = {2006}
}

@article{bentkus1986,
  title={Dependence of the Berry-Esseen estimate on the dimension},
  author={Bentkus, V},
  journal={Lithuanian Mathematical Journal},
  volume={26},
  number={2},
  pages={110-114},
  year={1986},
  publisher={Springer}
}

@article{bloznelis1989,
  title={On non-uniform estimate of convergence rate in multidimensional central limit theorem with stable limit law},
  author={Bloznelis, M},
  journal={Litov. Matem. Sb},
  volume={29},
  pages={350-365},
  year={1989}
}

@article{gotze1991,
  title={On the rate of convergence in the multivariate CLT},
  author={Gotze, F},
  journal={Annals of Probability},
  pages={724-739},
  volume={19},
  year={1991},
  publisher={JSTOR}
}

@article{nagaev2006,
  title={An estimate of the remainder term in the multidimensional central limit theorem},
  author={Nagaev, Sergey V},
  journal={Proceedings of the Third Japan—USSR Symposium on Probability Theory},
  volume={16},
  pages={419-438},
  year={2006}
}

@article{bentkus2003,
  title={On the dependence of the Berry-Esseen bound on dimension},
  author={Bentkus, Vidmantas},
  journal={Journal of Statistical Planning and Inference},
  volume={113},
  number={2},
  pages={385-402},
  year={2003},
  publisher={Elsevier}
}

@article{central,
  title={Central limit theorems for high dimensional dependent data},
  author={Chang, Jinyuan and Chen, Xiaohui and Wu, Mingcong},
  journal={Bernoulli},
  volume={30},
  number={1},
  pages={712-742},
  year={2024}
}
\bibliographystyle{rfs}
\end{document}